\documentclass[aps,prd,twocolumn,nofootinbib,showpacs,preprintnumbers]{revtex4-1}
\usepackage{natbib}
\usepackage[pdftex]{graphicx}
\usepackage[usenames]{color}
\usepackage[dvips]{epsfig}
\usepackage{youngtab}
\usepackage{hyperref}
\usepackage{bbm}
\usepackage{marvosym}
\usepackage{verbatim}
\usepackage{psfrag}
\usepackage{slashed}

\usepackage{amsmath}
\usepackage{amssymb}
\usepackage{wasysym}
\usepackage{multirow}
\usepackage{enumerate}
\usepackage{adjustbox}
\usepackage{mathtools}
\usepackage{diagbox}%used in tables to display slash in a box
\usepackage{array} % for '\extrarowheight' macro
\usepackage{amssymb}	% Mathe
\usepackage{booktabs,amsmath}
\usepackage{ bbold }
\usepackage[table,x11names]{xcolor}
\definecolor{maroon}{cmyk}{0,0.87,0.68,0.32}
\usepackage{amsfonts}

\definecolor{boxcolor}{HTML}{ffe6a1}

\def \sgn {\rm sgn}

\newcommand{\Wg}{{\rm W\hspace{-0.7mm}g}}
\newcommand{\tWg}{{\rm \tilde{W}\hspace{-0.7mm}g}}

\makeatletter
\newcommand{\raisemath}[1]{\mathpalette{\raisem@th{#1}}}
\newcommand{\raisem@th}[3]{\raisebox{#1}{$#2#3$}}

\newcommand{\ijkl}{_{i\,\raisemath{3pt}{j},k\,\raisemath{3pt}{l}}}

\newcommand{\beqn} {\begin{equation}}
\newcommand{\eqn} {\end{equation}}

\def \beq{\begin{equation}}
\def \eeq{\end{equation}}
\def \bea{\begin{eqnarray}}
\def \eea{\end{eqnarray}}

\def \Tr {{\rm Tr}}
\def \tr {{\rm tr}}

\def \bet0{\beta_0}
\def \bet1{\beta_1}
\def \simgt{\,\rlap{\lower 7.5 pt\hbox{$\mathchar \sim$}}\raise 3 pt \hbox{$>$}\,}
\def \simlt{\,\rlap{\lower 7.5 pt\hbox{$\mathchar \sim$}}\raise 3 pt \hbox{$<$}\,}

\def\lsim{\raise0.3ex\hbox{$<$\kern-0.75em\raise-1.1ex\hbox{$\sim$}}}
\def\gsim{\raise0.3ex\hbox{$>$\kern-0.75em\raise-1.1ex\hbox{$\sim$}}}

\newcommand{\SU}{{\rm SU}}
\newcommand{\U}{{\rm U}}

\newcommand{\rhoname}{DOI}
\newcommand{\rhonames}{DOIs}
\def \len {\rm len}

\DeclareMathAlphabet{\mathpzc}{OT1}{pzc}{m}{it}

\begin{document}
%\linenumbers

\title{A new dual representation for staggered lattice QCD}
\author{
G.~Gagliardi$^{\rm a}$,
W.~Unger$^{\rm a}$
}
\affiliation{$^{\rm a}$ Fakult\"at f\"ur Physik, Bielefeld University\\
}

\begin{abstract}
We propose a new strategy to evaluate the partition function of lattice QCD with Wilson gauge action coupled to staggered fermions, based on a strong coupling  expansion in the inverse bare gauge coupling $\beta= 2N/g^{2}$. Our method makes use of the recently developed formalism to evaluate the $\SU(N)$ $1-$link integrals and consists in an exact rewriting of the partition function in terms of a set of additional dual degrees of freedom which we call "Decoupling Operator Indices" (\rhoname{}). The method is not limited to any particular number of dimensions or gauge group $\U(N)$, $\SU(N)$. In terms of the \rhoname{} the system takes the form of a Tensor Network which can be simulated using Worm-like algorithms. Higher order $\beta$-corrections to strong coupling lattice QCD can be, in principle, systematically evaluated, helping to answer the question whether the finite density sign problem remains mild when plaquette contributions are included. Issues related to the complexity of the description and strategies for the stochastic evaluation of the partition function are discussed.
\end{abstract}

\pacs{02.20.−a, 02.30.Cj, 05.10.−a, 12.38.Gc, }
\maketitle
%\tableofcontents

\section{Introduction}
Lattice QCD at finite baryon density suffers from the notorious sign problem \cite{deForcrand2010}. 
In a nutshell, the numerical sign problem arises because the weights of the partition function are not positive definite, prohibiting importance sampling in Monte Carlo simulations. 
One of the several promising approaches to tackle the various sign problems in lattice field theories or spin systems are dual formulations. The basic idea is to rewrite the partition function by replacing the original (possibly continuous) degrees of freedom by new discrete degrees of freedom, such that the numerical sign problem of the new representation is milder or absent \cite{Gattringer2016b}.

The idea of dual representations is old, and in the last decade, many different sign problems have been solved in this way. 
Some of the hallmarks in the context of spin models are the O(N) and CP(N-1) models \cite{Wolff2009,Wolff2010,Bruckmann2015}, and in the context of lattice field theories are the charged scalar $\phi^4$ theory \cite{Gattringer2012}, the Abelian gauge-Higgs model \cite{Mercado2013b,Gattringer2018}, the SU(2) principle chiral model 
\cite{Gattringer2017} and scalar QCD \cite{Bruckmann2017}.
The term ``dual representation'' may seem as a misnomer (they are not duality transformations), but it has been established as an umbrella term for representations of specific type: the representations are obtained by integrating out the original degrees of freedom and by introducing discrete variables that encode nearest neighbour interaction, e.g. so-called bond variables. These are based on a high temperature or strong coupling expansion 
\cite{Wolff2008a,Wolff2008b} or similar Taylor expansions, and can be expressed in terms of oriented fluxes and/or unoriented occupation numbers (usually called monomers and dimers). A dual representation is then oftentimes called a world-line representation, or a dimerization, or is a combination of both. An important feature is that the original symmetries of the system are translated into constraints such as flux conservation or restrictions on the allowed occupation numbers. Typically these constraints are central in Monte Carlo simulations such as in the Worm algorithm \cite{Prokofev2001} or generalizations thereof \cite{Mercado2013}. Dual representations are in general not unique: a model can have several dual representations which may have different residual sign problems. 
In some cases, a dual representations can introduce a sign problem that did not exist in the original formulation. An important example is the lattice Schwinger model at finite quark mass. 

The focus of this paper is whether dual representations can be successfully applied to lattice QCD at finite baryon density, which has a severe sign problem in the usual representation, where fermions are integrated out, resulting in the fermion determinant.  
The standard approach is then Hybrid Monte Carlo. At finite baryon chemical potential $\mu_B$, the fermion determinant becomes complex, resulting in the sign (complex phase) problem.
Many strategies are available to circumvent the sign problem for small values of the chemical potential, like the Taylor expansion method~\cite{Allton2005}, the use of an imaginary chemical potential~\cite{deForcrand2003,DElia:2002tig} and reweighting~\cite{Fodor2001}. The latter has led to a first estimate of the position of the critical endpoint on a coarse lattice~\cite{Fodor2004}. In general, however, reweighting may suffer from the lack of overlap between the sampled $\mu_B=0$ ensemble and the target ensemble at $\mu_B>0$.
More recently, other approaches that are not limited to small $\mu_{B}$ have been proposed, such as the Complex Langevin approach~\cite{Berges:2005yt,Sexty2013}, the Lefschetz thimble approach \cite{Cristoforetti:2012su,Schmidt2017,DiRenzo2017}, or the density of states method \cite{Langfeld2012}.
To name also some approaches that are in the spirit of a dual representation: the 3-dimensional effective theory \cite{Fromm2012,Glesaaen2015} (a joint strong coupling and hopping parameter expansion that can be mapped to SU(3) spin model), decoupling the gauge links using Hubbard-Stratonovich transformations \cite{Vairinhos2015}, ``Induced QCD'' based on an alternative discretization of Yang Mills Theory \cite{Brandt2016,Brandt2019}. All these approaches have their shortcomings, and a method that allows to simulate lattice QCD at finite density has not yet been established.  

A dual representation of lattice QCD has only been derived in the strong coupling regime: the classical formulation in terms of a monomer-dimer-polymer  system has been both addressed via mean-field \cite{Kawamoto1981,KlubergStern1983,Faldt1985,Miura2016} and Monte Carlo \cite{Rossi1984,Karsch1989,Adams2003,Forcrand2010} and is valid only in the strong coupling limit.  More recently also the leading order gauge corrections have been included \cite{deForcrand2014,Gagliardi:2017uag}. 
At strong coupling, also the fermion bag approach has been used \cite{Marchis2018,Orasch2019} and continuous time methods have been applied \cite{Unger2011,Klegrewe2018}. 
Beyond the leading order, a dual formulation for lattice QCD is notoriously difficult. First attempts were made using a character expansion~\cite{Cherrington:2007ax, Cherrington:2009fx} and the so-called Abelian Colour Cycles~\cite{Gattringer2016a}. Our ultimate goal is to find a dual representation for lattice QCD: we propose a new approach based on a 
combined expansion of the Wilson plaquette action (strong coupling expansion) and of the staggered action (hopping and quark mass expansion) to all orders. 
The integration order is, as in the case of the strong coupling formulation, swapped, with the gauge integral being performed first while Grassmann integration is carried out after a reparameterization of the link integrals. % in term of \textit{generalized orthogonal projectors}. 
The strong coupling methods we use go back to the early days of lattice QCD, where computers for large scale simulations were not yet available \cite{Creutz:1978ub,Drouffe1983}.
But only due to recent progress in the computation of $1$-link integrals (invariant polynomial integration \cite{Gagliardi2018,Borisenko:2018csw}) we have complete control on the evaluation of the resulting Boltzmann weight ending up with a fully dualized partition function.
The challenge when going beyond the leading order correction is that this dual representation needs to capture non-local effects: it is no longer possible to write the partition function as product of site weights and bond weights only. The basic objects of our dual representation have a tensorial structure. In this paper we show a strategy to compute these tensors. 
Our method is not restricted to staggered fermions and can readily be applied to Wilson fermions as well.\\

The paper is organized as follows: In section~\ref{I} we review the computation of link integrals and introduce the $\SU(N)$ Decoupling Operators which constitutes the building blocks of the whole dualization process. In section~\ref{II} we sketch the steps needed to recover the color singlet Boltzmann weight from the computation of polynomial gauge integrals. In section ~\ref{III} the dualized partition function will be presented along with the expression of various observables in terms of the dual degrees of freedom and a discussion about the sign problem. In section ~\ref{IV} numerical crosschecks from exact enumeration in low dimensional systems will be shown. Finally in section ~\ref{V} we draw our conclusions.     

\section{Strong Coupling Expansion and Link Integration}
\label{I}
We consider the finite density partition function of lattice gauge theory with $\SU(N)$ gauge group, using the Wilson gauge action and one flavor of unrooted staggered fermions $\{\bar{\chi},\chi\}$ with lattice quark mass $\hat{m}_q=am_q$ 
\begin{equation}
\label{PartitionFunction}
\mathcal{Z}= \!\int\!\!\left[\mathcal{D}\bar{\chi}\chi\right]e^{-2\hat{m}_{q}\bar{\chi}_{x}\chi_{x}}\!\left[\prod_{\ell}\int_{\SU(N)}\!\!\!\!\! DU_{\ell}\right]e^{-S_{g}[U]-D_{f}[\bar{\chi},\chi,U]},
\end{equation} \\
where $\ell=(x,\mu)$ and $x$ stand respectively for lattice links and sites and $DU$ is the Haar measure. The gauge links $U_{\ell}$ are $\SU(N)$ elements, while $S_{g}$ and $D_{f}$ are respectively the plaquette action and the massless staggered Dirac operator:
\begin{widetext}
\begin{align}
S_{g}[U] &= -\frac{\beta}{2N}\sum_{x,\mu<\nu}\Tr \, U_{x,\mu}U_{x+\mu,\nu}U^{\dag}_{x+\nu,\mu}U^{\dag}_{x,\nu} + h.c. = -\frac{\beta}{2N}\sum_{p} \Tr U_{p} + \Tr U^{\dag}_{p}, \nonumber\\
%
%D_{f}[\bar{\chi},\chi,U] &= \sum_{x,\mu} \eta_{\mu}(x)e^{+\mu_{q}\delta_{\mu,0}}\bar{\chi}_{x}U_{x,\mu}\chi_{x+\mu} \nonumber \\
%&\phantom{=}-\sum_{x,\mu} \eta_{\mu}(x)e^{-\mu_{q}\delta_{\mu,0}}\bar{\chi}_{x+\mu}U^{\dag}_{x,\mu}\chi_{x} \nonumber \\
%
D_{f}[\bar{\chi},\chi,U] &= \sum_{x,\mu} \eta_{\mu}(x)\left( e^{+\mu_{q}\delta_{\mu,0}}\bar{\chi}_{x}U_{x,\mu}\chi_{x+\mu}
-e^{-\mu_{q}\delta_{\mu,0}}\bar{\chi}_{x+\mu}U^{\dag}_{x,\mu}\chi_{x}
\right)
\equiv \sum_{\ell} \Tr U_{\ell}\mathcal{M}^{\dag}_{\ell} + \Tr U^{\dag}_{\ell}\mathcal{M}_{\ell},
\end{align}  
\end{widetext}
where $\mu_{q}=\frac{1}{N}\mu_B$ is the lattice quark chemical potential and $\eta_{\mu}(x)$ are the usual staggered phases. All traces are intended to be over color indices and in the following we will always use the letter $p$ to label lattice plaquettes. \\\\
The first step in the dualization process is to perform a combined Taylor expansion of Eq.~(\ref{PartitionFunction}) in the reduced gauge coupling $\tilde{\beta} \equiv \frac{\beta}{2N}=\frac{1}{g^{2}}$  and quark mass $\hat{m}_{q}$:
\begin{widetext}
\begin{align}
\mathcal{Z}(\beta,\hat{m}_q)&=
{\displaystyle \sum_{\substack{\{n_{p},\bar{n}_{p}\} \\ \{d_{\ell},\bar{d}_{\ell},m_{x} \}  }}}\!\!\!\prod_{p}\frac{\tilde{\beta}^{n_{p}+\bar{n}_{p}}}{n_{p}!\bar{n}_{p}!}\prod_{\ell}\frac{1}{d_{\ell}!\bar{d}_{\ell}!}{\displaystyle \prod_{x}}\frac{(2\hat{m}_{q})^{m_{x}}}{m_{x}!}\,\boldsymbol{\mathcal{G}}_{n_{p},\bar{n}_{p},d_{\ell},\bar{d}_{\ell},m_{x}},
\label{SCE}
\\
\boldsymbol{\mathcal{G}}_{n_{p},\bar{n}_{p},d_{\ell},\bar{d}_{\ell},m_{x}} &= \int\mathcal{D}\left[\chi\bar{\chi}\right](\bar{\chi}_{x}\chi_{x})^{m_{x}}{\displaystyle \prod_{\ell,p}}\int DU_{\ell}\Tr[U_{p}]^{n_{p}}\Tr[U^{\dag}_{p}]^{\bar{n}_{p}}\Tr[U_{\ell}\mathcal{M}^{\dag}_{\ell}]^{d_{\ell}}\Tr[U^{\dag}_{\ell}\mathcal{M}_{\ell}]^{\bar{d}_{\ell}}.
\label{G}
\end{align} 
\end{widetext}
The sum is over the positive integers that single out a particular term in the expansion: ($\bar{n}_{p}$) $n_{p}$ is called the (anti-) plaquette occupation number, $m_{x}$ the monomer number and $d_{\ell}$, $\bar{d}_{\ell}$ stem from the expansion of the massless staggered Dirac operator in forward ($d_{\ell}$) and backward ($\bar{d}_{\ell}$) direction. 
The quantity $\boldsymbol{\mathcal{G}}$ contains the non-local part of the computation and is given by a Gauge$+$Grassmann integral over the whole lattice.\\ 

Our dualization corresponds to exactly integrate out the gauge links $U_{x,\mu}$ and 
the Grassmann field $\bar{\chi},\chi$, trading the original degrees of freedom with the integer variables appearing in Eq.~(\ref{SCE}) . This can be achieved by splitting the computation of $\mathcal{G}$ in two steps:
\begin{enumerate}[1)]
\item The traces appearing in Eq.~(\ref{G}) are written explicitly: we do not perform the matrix multiplication, leaving the color indices uncontracted. As a consequence, the gauge integral $\prod_{\ell}\int_{\SU(N)}DU_{\ell}$, becomes a disjoint product of monomial integrals with open color indices and we integrate out  every gauge link independently. 
\item After gauge integration, some of the open color indices need to be contracted between links that share a common site such that the plaquette terms are recovered. The remaining indices are contracted with the Grassmann-integrated quark fields. We postpone the description of this step to Section~\ref{II}.    
\end{enumerate} 
If the matrix multiplications are not performed, the link integrals to be computed assume the following general form:
\begin{equation}
\label{I-integral}
\mathcal{I}^{a,b}\ijkl = \int_{\SU(N)}DU\,\, U_{i_{1}}^{\,j_{1}} \cdots U_{i_{a}}^{\,j_{a}}U^{\dag \,l_{1}}_{k_{1}} \cdots U^{\dag \,l_{b}}_{k_{b}},
\end{equation}
where the values $a$, $b$ depend on the dual degrees of freedom $\{n_{p},\bar{n}_{p},d_{\ell},\bar{d}_{\ell}\}$ and we make use of the multi-index notation:
\begin{align}
i&=(i_1,i_2, \ldots, i_a),& 
j&=(j_1,j_2, \ldots, j_a),\nonumber\\
k&=(k_1,k_2, \ldots, k_b),&
l&=(l_1,l_2, \ldots, l_b).
\end{align}
Due to the properties of the $\SU(N)$ invariant Haar measure, the integrals in Eq.~(\ref{I-integral}) are non-zero only when the difference $a-b$ is an integer multiple of $N$. As it will be explained in the next section, this corresponds to a (gauge-) constraint for the dual degrees of freedom. We define:
\begin{equation}
q=\frac{|a-b|}{N} ,\qquad q \in \mathbbm{N},
\end{equation}
and for $\U(N)$ gauge theory $q=0$.
Invariant integration over compact groups have been studied extensively in the last decades~\cite{Weingarten:1977ya,Drouffe1983,Creutz:1978ub,Bars:1980yy,Bars:1979xb,Eriksson:1980rq,BROWER1981699,Carlsson:2008dh,Collins1,Collins2,Collins3,novak2008complete,novaes2014elementary,Zuber_2016}. Although many results concerning the $\U(N)$ group are known since many years, only recently the $\SU(N)$ generalization has been found~\cite{Borisenko:2018csw,Gagliardi2018}. Integrals of the type Eq.~(\ref{I-integral}) are now known in closed form in term of  \textit{Generalized Weingarten Functions}. The interested reader will find our derivation in Appendix~\ref{App1}. Here we only quote the main result assuming, without loss of generality, $a>b$ ($a=qN+p$, $b=p$):
\begin{equation}
\label{I-integral-computation}
\mathcal{I}^{qN+p,p}\ijkl \propto\!\!\!{\displaystyle \sum_{(\alpha,\beta)}}{\displaystyle \sum_{\pi,\sigma \in S_{p}}}\epsilon^{\otimes q}_{i_{\{\alpha\}}}\delta_{i_{\{\beta\}}}^{l_{\pi}}\tWg_{N}^{q,p}(\pi\circ\sigma^{-1})\epsilon^{\otimes q, j_{\{\alpha\}}}\delta_{k_{\sigma}}^{\;j_{\{\beta\}}}.
\end{equation}
In the previous equation, $\epsilon^{\otimes q}$ is a shortcut for the $q-$fold product of Levi-Civita epsilon tensors and $\delta_{i}^{\;l_{\pi}}$, $\delta_{k_{\sigma}}^{\;j}$ are the generalized Kronecker deltas where the indices are reordered according to the permutations $\pi$ and $\sigma$. The leftmost sum with:
\begin{align}
\label{partitions}
\alpha &= \{\alpha_{1},\ldots,\alpha_{q}\},& |\alpha_{r}| &= N,& |\beta | &= p
\end{align} is carried over the $\frac{(qN+p)!}{q!N!^{q}p!}$ possible ways of partitioning the color indices $i$, $j$ (which are $qN+p$) into the $q$ epsilon tensors of size $N$ and into the delta function of size $p$. All the partitions obtained from each other by only permuting the $\alpha_{r}$ in Eq.~(\ref{partitions}) are equivalent. Also, note that in Eq.~(\ref{I-integral-computation}) the $i$ and $j$ indices are partitioned in the same way.
As in the $\U(N)$ case, a further summation over all possible permutation of indices in the delta functions (sum over $\pi, \sigma$) is present. Every term in the double sum is weighted by the function $\tWg^{q,p}_{N}$, which is a class function of the symmetric group $S_{p}$ and represents the natural generalization of the Weingarten functions \nolinebreak{$\Wg^{p}_{N}=\tWg^{0,p}_{N}$} appearing in the $\U(N)$ result~\cite{Collins1,Collins2}. Their expression in terms of the characters $\chi^{\lambda}$ of the symmetric group is:
\begin{align}
\label{Weingarten}
\tWg_{N}^{q,p}(\pi) &= \sum\limits_{\substack{\lambda \vdash p \\ \len(\lambda) \leq N}}\frac{1}{(p!)^{2}}\frac{f_{\lambda}^{2}\chi^{\lambda}(\pi)}{D_{\lambda,N+q}}, \nonumber \\
\lambda\vdash p &\equiv \left\{ \lambda_{1} \geq \ldots \geq \lambda_{\len(\lambda)} > 0 \, \left| \,\, {\displaystyle \sum_{i=0}^{\ell(\lambda)}}\lambda_{i}=p \right\}\right..
\end{align} 
The sum is over the irreducible representations (irreps)\footnote{The $S_{n}$ irreps are in 1-1 correspondence with the integer partitions of $n$. In Eq.~(\ref{Weingarten}) only the irreps that correspond to integer partitions with at most $N$ parts contribute.} of the symmetric group $S_{p}$, while $f_{\lambda}$ is the dimension  of the irrep $\lambda$ of $S_{p}$ and $D_{\lambda,N+q}$ is the dimension of the U($N+q$) irrep with highest weight $\{\lambda_{1},\ldots,\lambda_{\len(\lambda)},0,\ldots\}$. \\

By inspecting Eq.~(\ref{I-integral-computation}), it seems tempting to consider the permutations $\pi,\sigma$ as an additional degree of freedom to be evaluated stochastically and to proceed with the index contraction considering single terms in the sum of Eq.~(\ref{I-integral-computation}). Unfortunately, the sign of the generalized Weingarten functions strongly oscillates, preventing the application of standard Monte Carlo methods. A similar failure is observed in~\cite{Cherrington:2009fx} where the role of the $\tWg$'s is played by the $18j$ symbols. 
Instead, we found it useful to exploit the knowledge of the character expansion in Eq.~(\ref{Weingarten}) to reparameterize the $\mathcal{I}$-integral. 
As a starting point we write the $S_{p}$ characters as a matrix product of the corresponding matrix representation:
\begin{align}
\chi^{\lambda}(\pi\circ\sigma^{-1}) &\equiv \Tr\left( M^{\lambda}(\pi)M^{\lambda}(\sigma^{-1})\right).
\end{align}
Writing the matrix product explicitly, we are able to cast the Weingarten functions and (after summing over the permutations) the $\mathcal{I}$-integrals in the following form:
\begin{widetext}
\begin{align}
\tWg_{N}^{q,p}(\pi\circ\sigma^{-1})&=
 %\sum\limits_{\substack{\lambda \vdash p \\ \ell(\lambda) \leq N}}\frac{1}{(p!)^{2}}\frac{f_{\lambda}^{2}\chi^{\lambda}(\pi\circ\sigma^ {-1})}%%{D_{\lambda,N+q}}=
\sum\limits_{\substack{\lambda \vdash p \\ \ell(\lambda) \leq N}}\sum_{m,n=1}^{f_{\lambda}}\!\left(\frac{1}{p!}\frac{f_{\lambda}}{\sqrt{D_{\lambda,N+q}}}M^{\lambda}_{mn}(\pi)\!\right)\left(\frac{1}{p!}\frac{f_{\lambda}}{\sqrt{D_{\lambda,N+q}}}M^{\lambda}_{mn}(\sigma)\!\right), \label{Weingarten_expansion} \\
\mathcal{I}^{qN+p,p}\ijkl &\propto {\displaystyle \sum_{(\alpha,\beta)}}\sum\limits_{\substack{\lambda \vdash p \\ \ell(\lambda) \leq N}}\sum_{m,n=1}^{f_{\lambda}}\!\!\left({\displaystyle \sum_{\pi}}\frac{1}{p!}\frac{f_{\lambda}}{\sqrt{D_{\lambda,N+q}}}M^{\lambda}_{mn}(\pi)\epsilon^{\otimes q}_{i_{\{\alpha\}}}\delta_{i_{\{\beta\}}}^{\;l_{\pi}}\!\right)\left({\displaystyle \sum_{\sigma}}\frac{1}{p!}\frac{f_{\lambda}}{\sqrt{D_{\lambda,N+q}}}M^{\lambda}_{mn}(\sigma)\epsilon^{\otimes q, j_{\{\alpha\}}}\delta_{k_{\sigma}}^{\;j_{\{\beta\}}}\right), \label{I-integral_expansion}
\end{align} 
\end{widetext}
where the matrices $M^{\lambda}$ have been chosen to be orthogonal.\footnote{Every finite group admits a unitary irrep. In the case of the symmetric group the matrix elements can be also chosen to be real. This basis is known as the Young's orthogonal form.}
The quantities in the brackets of Eq.~(\ref{I-integral_expansion}) generalize the $\U(N)$ Orthogonal Operators~\cite{Digest,Christensen:2014dia} (where the summation over $(\alpha,\beta)$ and the epsilon tensors are absent), and represent the building blocks of our dualization. The orthogonality property does not generalize to the $\SU(N)$ case, hence we will refer to them as the \textit{$\SU(N)$ Decoupling Operators}. They are identified by a given partition $(\alpha,\beta)$ and by choosing a given matrix element $(m,n)$ of an irrep $\lambda$ of $S_{p}$. We denote the latter as $(m,n)_{\lambda}$. Moreover, to make the expression in Eq.~(\ref{I-integral_expansion}) more compact, we collect $(\alpha,\beta)$ and $(m,n)_{\lambda}$ into a multi-index $\rho = [(\alpha,\beta), (m,n)_{\lambda}]$ so that we can write the $\SU(N)$ Decoupling Operators $P^{\rho}$ and the $\mathcal{I}$-integrals as:
\begin{align}
\left(P^{\rho}\right)^{\;l}_{i} &= {\displaystyle \sum_{\pi}}\frac{1}{p!}\frac{f_{\lambda}}{\sqrt{D_{\lambda,N+q}}}M^{\lambda}_{mn}(\pi)\epsilon^{\otimes q}_{i_{\{\alpha\}}}\delta_{i_{\{\beta\}}}^{\;l_{\pi}}, \label{DecouplingOperators} \\
\mathcal{I}^{qN+p,p}\ijkl &= {\displaystyle \prod_{r=1}^{N-1}}\frac{r!}{(r+q)!}\sum_{\rho}\left(P^{\rho}\right)^{\;l}_{i}\left(P^{\rho}\right)^{\;j}_{k},  \label{Final-I-Integral} \\
n_{\rho}  &= \frac{(qN+p)!}{q!N!^{q}p!} {\displaystyle \sum_{\lambda\vdash p}^{\ell(\lambda) \leq N}}f_{\lambda}^{2}, \label{numberOfOperators}
\end{align}
where $n_{\rho}$ is the total number of operator indices.
The operators $P^{\rho}$ decouple the colour indices $i, l$ and $k, j$ in the $\mathcal{I}$-integral and its computation has been automatized by using the standard \textit{hook rule} to determine $f^{\lambda}$ and $D_{\lambda, N}$. The irreducible matrix elements $M^{\lambda}_{mn}(\pi)$ in the orthogonal representation are computed numerically decomposing the permutation $\pi$ as a product of adjacent transpositions $\tau^{j,j+1}$ and then using the axial distance formula to compute the matrix representation associated to them (see~\cite{Digest} p. 8). The quantity $\rho$, which identifies a given operator in Eq.~(\ref{Final-I-Integral}), will play an important role in the following. We will refer to it as \textit{Decoupling Operator Index} (\rhoname), which can be cast into an integer in the range $\{1,\ldots,n_{\rho}\}$. 

\section{Index Contraction and Tensor Network}

\label{II}
\begin{figure*}
\begin{center}
\includegraphics[scale=0.20]{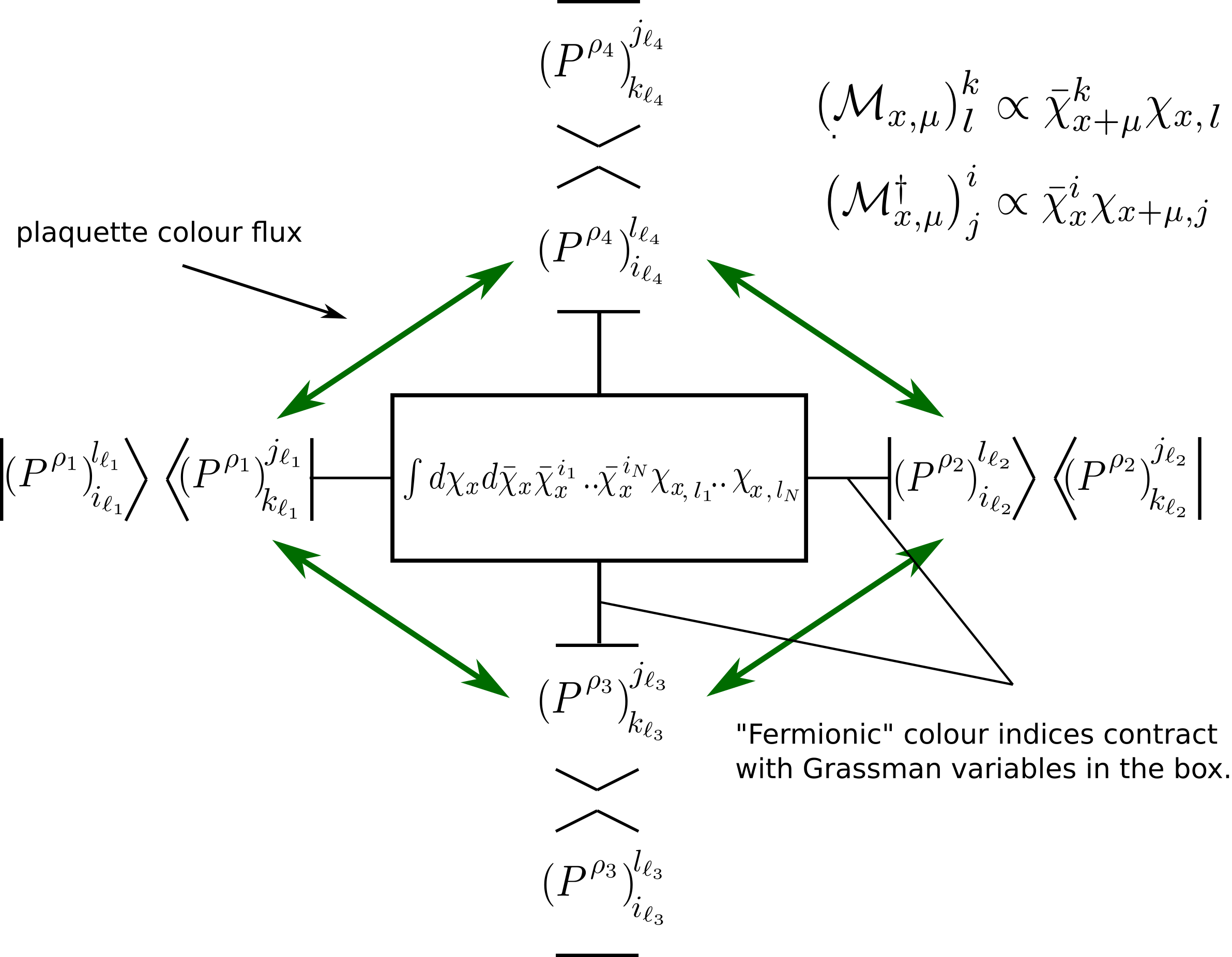}
\caption{Illustration of the contraction step in two dimensions: On each of the 4 links attached to the central lattice site the \rhonames{} $\{\rho_{1},\rho_{2},\rho_{3},\rho_{4}\}$ have been fixed. Decoupling Operators on the same link, undergo a disjoined contraction at two different lattice sites. The bra-ket notation only serves to display this feature. At any lattice site (e.g. the central box) the color indices of the 4 operators are completely saturated. Depending on the plaquette and anti-plaquette occupation number $\{n_{p},\bar{n}_{p}\}$ on the 4 plaquettes attached to the site, some of the color indices are contracted between the operators (green arrows). Instead, the color indices stemming from the hopping expansion of the staggered action are contracted with the reordered Grassmann variables at site $x$. The result is a scalar quantity which only depends on the value of the \rhonames{} and on the dual degrees of freedom $\{n_{p},\bar{n}_{p},d_{\ell},\bar{d}_{\ell},m_{x}\}$.   }
\label{Fig1}
\end{center}
\end{figure*}

Given the result in Eq.~(\ref{Final-I-Integral}), the next step is to perform the contraction of the color indices $\{i,j,k,l\}$ in the $\mathcal{I}$-integrals making use of decomposition in terms of the operators $P^{\rho}$ obtained in the previous section.
This contraction must be performed for every lattice link $\ell$.  The (anti-) plaquette occupation numbers ($\bar{n}_{p}$) $n_{p}$ together with $d_{\ell}$, $\bar{d}_{\ell}$ determine how the contraction has to be performed in order to recover Eq.~(\ref{G}). We distinguish two types of color indices: those stemming from the expansion of the hopping term and those arising from the expansion of the Wilson gauge action. We will refer to them as the "\textit{fermionic}" and "\textit{gluonic}" color indices. The contraction rules for the fermionic color indices are uniquely determined by $d_{\ell}$ and $\bar{d}_{\ell}$. These indices are contracted with the Grassmann fields appearing in the corresponding fermionic matrices $\mathcal{M}_{\ell}$ and $\mathcal{M}^{\dag}_{\ell}$. Due to the nilpotency of the Grassmann measure, exactly $N$ (for $\SU(N)$) (anti-) fermion fields ($\bar{\chi}_{x}$) $\chi_{x}$ have to be present at each site $x$ in order to obtain a non-zero contribution. This property will corresponds to a constraint on the allowed degrees of freedom $d_{\ell},\bar{d}_{\ell},m_{x}$. Similarly, gluonic color indices are contracted according to the plaquette they correspond to. In this case the contraction takes place between the color indices of the $\mathcal{I}$-integrals corresponding to links sharing a common site. The contraction rules are determined by the (anti-) plaquette occupation numbers and allows us to recover the plaquette terms in Eq.~(\ref{G}). \\

\begin{center}
\begin{figure}
\includegraphics[scale=0.28]{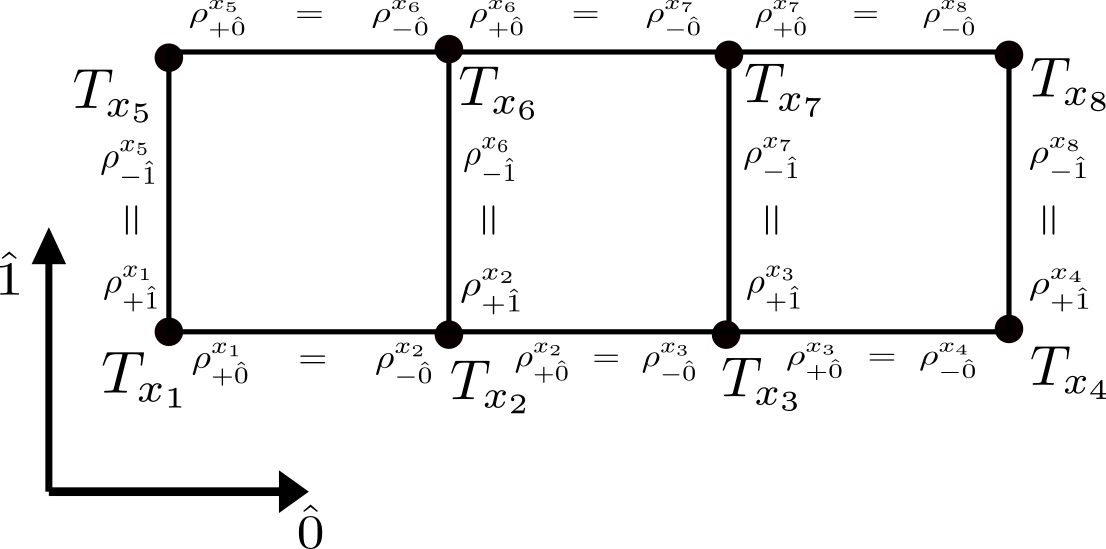}
\caption{The Tensor Network resulting from the dual description: Depending on the dual degrees of freedom at any lattice site the tensor $T_{x}$ is evaluated. Given two neighbouring sites $x$ and $x+\mu$, the tensor index on the common link is contracted $(\rho^{x}_{\mu}=\rho^{x+\mu}_{-\mu})$. The value of $\boldsymbol{\mathcal{G}}$ is the scalar quantity obtained by contracting all pairs of indices between lattice neighbours. In the figure, the tensor indices have been displaced for visualization purposes. }
\label{Fig2}
\end{figure}
\end{center}

The key insight is that fixing the values of the \rhoname{} $\rho_{\ell}$ for each link $\ell$, makes the contraction step local. This means that the contraction of the color indices can be carried out independently at different lattice sites. In Fig.~\ref{Fig1} we illustrate the procedure in $d=2$. The extension to any number of dimensions is straightforward.
To see why the contraction at different lattice sites decouples, let us first consider the case of the gluonic color indices, and rewrite explicitly the definition of the plaquette and anti-plaquette: 
For any gauge link $U_\ell$ with $\ell=(x,\mu)$, the contribution from the product of traces $\Tr U_{p}, \Tr U_{p}^{\dagger}$ for all plaquettes $p$ containing the link $\ell$ can be gathered into products of matrix elements of $U_{\ell}$ and $U_{\ell}^{\dagger}$: 
\begin{widetext}
\begin{align}%
\label{Plaquette-explicit}
\Tr U_{p} =\big(U_{1}\big)_{i_{1}}^{\;j_{1}}\big(U_{2}\big)_{i_{2}'}^{\;j_{2}'}\big(U_{3}^{\dag}\big)_{k_ {3}'}^{\;l_{3}'}\big(U_{4}^{\dag}\big)_{k_ {4}'}^{\;l_{4}'} \,\delta_{j_{1}}^{\;i_{2}'}\,\delta_{j_{2}'}^{\;k_{3}'}\,\delta_{l_{3}'}^{\;k_{4}'}\,\delta_{l_{4}'}^{\;i_{1}},\qquad
\Tr U_{p}^{\dag} = \big(U_{1}^{\dag}\big)_{k_ {1}}^{\;l_{1}}\big(U_{2}^{\dag}\big)_{k_ {2}'}^{\;l_{2}'}\big(U_{3}\big)_{i_ {3}'}^{\;j_{3}'}\big(U_{4}\big)_{i_{4}'}^{\;j_{4}'} \,\delta_{l_{2}'}^{\;k_{1}}\,\delta_{j_{3}'}^{\;k_{2}'}\,\delta_{j_{4}'}^{\;i_{3}'}\,\delta^{\;l_{1}}_{i_{4}'},\\
\prod_{\{p=(x,\mu,\nu)\,\mid\, \ell\in p\}}^{n} (U_\ell)_{i_1}^{\;j_1} \ldots (U_\ell)_{i_{n_p}}^{\;j_{n_p}}\;\; (U_\ell^\dagger)_{k_1}^{\;l_1} \ldots (U_\ell^\dagger)_{k_{\bar{n}_p}}^{\;l_{\bar{n}_p}}\quad\hookrightarrow\quad \prod_{\{p=(x,\mu,\nu)\,\mid\, \ell\in p\}}^{n}(\Tr U_{p})^{n_{p}} (\Tr U_{p}^\dagger)^{\bar{n}_{p}}, 
\label{ExpandTrace}
\end{align}
\end{widetext}
where $U_{1},\ldots, U_{4}$ are the four links contained in the plaquette and summation over repeating color indices is implied.
The l.h.s.~of Eq.~(\ref{ExpandTrace}) thus contributes to the gluonic color indices $\{i,l\}, \{k,j\}$ within $\mathcal{I}\ijkl$. Therefore, given the structure of the operators in Eq.~(\ref{Final-I-Integral}), $\left(P^{\rho}\right)^{\;l}_{i}$ and $\left(P^{\rho}\right)^{\;k}_{j}$ contract respectively with the operators attached to site $x$ and $x+\mu$.
Fermionic color indices arise instead from terms of the form:
\begin{align}
\Tr[U_{\ell}\mathcal{M}^{\dag}_{\ell}]^{d_{\ell}},&\quad\Tr[U^{\dag}_{\ell}\mathcal{M}_{\ell}]^{\bar{d}_{\ell}}.
\end{align}  
They can be written explicitly as $\big(\ell=(x,\mu)\big)$\footnote{The dependency of $\mathcal{M},\mathcal{M}^{\dag}$ on $\eta_{\mu}$ and $\mu_{q}$ can be factored out as it will be shown in the dual partition function Eq.~(\ref{DualizedPartitionFunction}). }:
\begin{align}
\Tr[U_{\ell}\mathcal{M}^{\dag}_{\ell}]^{d_{\ell}}&\propto {\displaystyle \prod_{a=1}^{d_{\ell}}}\left(U_{x,\mu}\right)_{i_{a}}^{\,j_{a}}\bar{\chi}_{x}^{i_{a}}\chi_{x+\mu, j_{a}}, \nonumber \\ 
\Tr[U^{\dag}_{\ell}\mathcal{M}_{\ell}]^{\bar{d}_{\ell}} &\propto {\displaystyle \prod_{b=1}^{\bar{d}_{\ell}}}\left(U_{x,\mu}^{\dag}\right)_{k_{b}}^{\,l_{b}}\bar{\chi}_{x+\mu}^{k_{b}}\chi_{x, l_{b}},
\end{align}
and again by inspecting at the index structure of Eq.~(\ref{Final-I-Integral}) the indices $\{i,l\}$ of the first operator $\left(P^{\rho}\right)^{\;l}_{i}$  are contracted with the Grassmann variables at site $x$ while the indices $\{k,j\}$ of the second operator  with the Grassmann variables at site $x+\mu$. This concludes the proof of the locality of the contractions which is schematically shown in Fig.~\ref{Fig1}. At each site, the corresponding Grassmann integral is replaced with the usual product of two epsilon tensors:
\begin{equation}
\label{EpsilonTensor}
\int \left[d\bar{\chi}_{x}d\chi_{x}\right]\bar{\chi}_{x}^{i_{1}}\cdots\bar{\chi}_{x}^{i_{N}}\chi_{x, l_{1}}\cdots\chi_{x, l_{N}}= \epsilon^{i_{1}\cdots i_{N}}\epsilon_{l_{1}\cdots l_{N}}.
\end{equation}
Hence, on a $d$-dimensional hypercubic lattice, the  $2d$ operators attached to $x$, together with the epsilon tensors in Eq.~(\ref{EpsilonTensor}) are jointly contracted according to the values of the dual degrees of freedom. This gives a scalar quantity which only depends on the underlying dual degrees of freedom and on the values of the \rhonames{} on the links attached to $x$. The dependency on the dual degrees of freedom $\{n_{p},\bar{n}_{p},d_{\ell},\bar{d}_{\ell},m_{x}\}$ is local, in the sense that the contraction at site $x$ is completely determined by the monomer number $m_{x}$, by the values of $d_{\ell}$ and $\bar{d}_{\ell}$ on the $2d$ links attached to $x$, and on the (anti-) plaquette occupation numbers of the $2d(d-1)$ plaquettes attached to $x$. Different sites communicate only via the common \rhoname{}  $\rho$ on the shared leg. We can collect the scalar quantities obtained from the contraction of different Decoupling Operators in a tensor
\begin{align} 
\label{TensorDefinition}
T_{x}^{\rho^{x}_{-d}\cdots\rho^{x}_{d}}\left(\mathcal{D}_{x}\right)&\equiv \Tr_{\mathcal{D}_{x}}\left[{\displaystyle \prod_{\pm \mu}}P^{\rho^{x}_{\mu}}\right], \nonumber \\
\mathcal{D}_{x}&= \left\{m_{x},d_{x,\pm\mu}, n_{x,\mu\nu},\bar{n}_{x,\mu\nu}\right\},
\end{align}
where $\Tr_{\mathcal{D}_{x}}$ is a "reordered" trace in color space that depends on the local dual degrees of freedom $\mathcal{D}_{x}$ and tells us how to contract the color indices of the operators $P^{\rho^{x}_{\mu}}$ according to the rules discussed above.
In Eq.~\ref{TensorDefinition} the \rhonames{} depend on $\mathcal{D}_{x}$ implicitly due to the fact that the dual degrees of freedom determine the value of $(q,p)$ in the $\mathcal{I}$-integral Eqs.~(\ref{I-integral}),~(\ref{I-integral-computation}). The tensor elements $T_{x}^{\rho}$, can be computed numerically by building up the operators $P^{\rho^{x}_{\mu}}$ and saturating their color indices according to the contraction rules from $\mathcal{D}_{x}$. Given $T_{x}^{\rho}$, the value of $\boldsymbol{\mathcal{G}}$ is given, up to a global fermionic sign (see Sec.~\ref{III}), by
\begin{align}
\boldsymbol{\mathcal{G}}_{n_{p},\bar{n}_{p},d_{\ell},\bar{d}_{\ell},m_{x}} &= {\displaystyle \sum_{\{\rho^{x}_{\pm\mu}\big| \rho^{x}_{\mu} = \rho_{-\mu}^{x+\mu}\}}}{\displaystyle \prod_{x}}T_{x}^{\rho^{x}_{-d},\dots,\rho_{d}^{x}}(\mathcal{D}_{x}) ,
\end{align} \\   
and the constraint $\rho^{x}_{\mu} = \rho_{-\mu}^{x+\mu}$ just stems from the fact that \rhonames{} on the same link have to be equal as depicted in Fig~\ref{Fig2}. In this form, the system is represented by a Tensor Network where the value of $\boldsymbol{\mathcal{G}}$ is obtained by contracting the network to a scalar. 

In some cases the contraction of different operators produces the same tensor elements. For instance, two operators with \rhonames{} $[(\alpha,\beta),(m,n)_{\lambda}]$ and $[(\alpha',\beta'),(m,n)_{\lambda}]$ where $(\alpha,\beta)$ and $(\alpha',\beta')$ only differ by a permutation of fermionic color indices, will produce the same element up to a sign factor. This is clear since the fermionic color indices are always contracted with the Grassmann variables, and a permutation of fermionic color indices only amounts to a reordering of the corresponding indices in the epsilon tensors in Eq.~(\ref{EpsilonTensor}). The possible relative minus sign is however unimportant. In fact, it will always cancel when considering the contraction of the operator with same \rhoname{} and which lives on the same link. We therefore identify these \rhonames{} taking into account the combinatorial factor from their multiplicity. This reduces the size of the tensor $T_{x}$, hence the numerical cost of contracting the network.\\

As we already mentioned, not all sets of dual degrees of freedom are allowed. On each lattice link they have to combine in a way that the corresponding $\mathcal{I}$-integral is non-zero, while at any site exactly $N$ (anti-) fermions carrying different colors must be present. We refer to these two constraints as Gauge and Grassmann constraint. Introducing
\begin{align}
k_{\ell} &= \min\left\{ d_{\ell},\bar{d}_{\ell} \right\},& f_{\ell} &= d_{\ell} - \bar{d}_{\ell},
\end{align} \\
where $k_{\ell}$ is the \textit{dimer number} and $f_{\ell}$ the \textit{quark flux}, for each link $\ell = (x,\mu)$ the gauge constraint reads:
\begin{align}
\label{GaugeConstraint}
f_{x,\mu}&+{\displaystyle \sum_{\nu>\mu}}\bigg[\delta n_{\mu,\nu}(x)-\delta n_{\mu,\nu}(x-\nu)\bigg]  \nonumber \\
&- {\displaystyle \sum_{\nu < \mu}}\bigg[ \mu\leftrightarrow\nu\bigg] = N q_{x,\mu} ,\qquad
q_{x,\mu}\in\mathbb{Z},
\end{align} \\
where $\delta n_{\mu,\nu}(x)\equiv\delta n_{p} = n_{p} -\bar{n}_{p}$. For each site $x$, the Grassmann constraint requires in addition:
\begin{align}
\label{GrassmanConstraint}
&m_{x}+ \sum_{\pm \mu}\left( k_{x,\mu} +\frac{|f_{x,\mu}|}{2}\right) = N, \quad {\displaystyle \sum_{\pm \mu}} f_{x,\mu} = 0.
\end{align}
The Eqs.~(\ref{GaugeConstraint}),~(\ref{GrassmanConstraint}) generalize the constraint in the strong coupling limit (where $n_{p}=\bar{n}_{p}=0$ and $f_{x,\mu}= \pm N,0$\nolinebreak). Notice that in contrast to strong coupling QCD, dimers $(d_{x,\mu} \neq 0)$ and fluxes $(f_{x,\mu} \neq 0)$ are not mutually exclusive on a given link.  
The set $\{n_{p},\bar{n}_{p},f_{\ell},{k}_{\ell},m_{x}\}$ subject to Eqs.~(\ref{GaugeConstraint}),~(\ref{GrassmanConstraint}) along with the corresponding \rhonames{} define our final dual partition function.

\section{Partition Function in the Dual Representation}
\label{III}
\subsection{General properties}
\begin{center}
\begin{figure}
\includegraphics[scale=1.1]{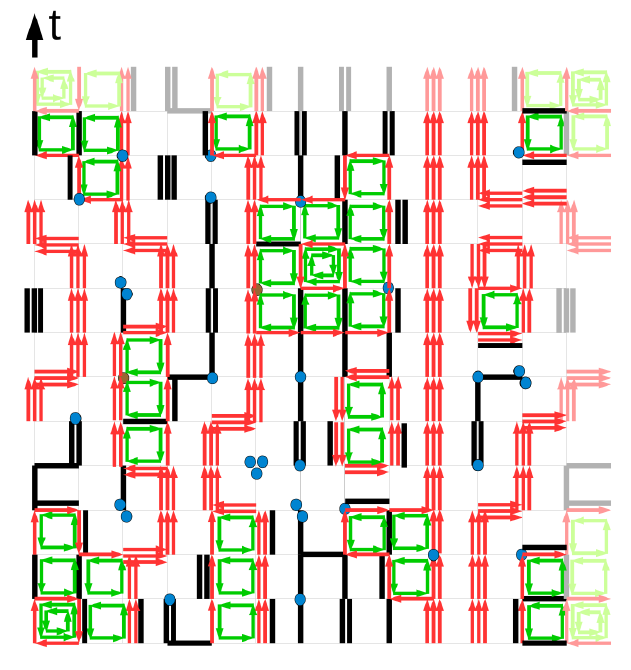}
\label{Fig3}
\caption{An allowed configuration in $d=2$ for $\SU(3)$: For each plaquette, a (counter- ) clockwise loop corresponds to one unit of ($n_{p} $) $\bar{n}_{p}$. On each site the monomer number $m_{x}$ is given by the number of circles, while on each bond the unoriented lines represent dimers (n lines for $k_{x,\mu}=n$). Every arrow represents instead one unit of flux $f_{x,\mu}$. The Grassmann constraint, in agreement with Eq.~(\ref{GrassmanConstraint}), is satisfied at each site with the net flux $\sum_{\mu}f_{x,\mu}$ being always zero. For every link, the difference between the total flux (gluons $+$ quarks) in positive and negative direction is a multiple of $N=3$. P.B.C. are employed.}
\end{figure}
\end{center}

Using the quantities defined in the previous sections, the partition function Eq.~(\ref{PartitionFunction}) can be rewritten as:
\begin{widetext}
\begin{equation}
\label{DualizedPartitionFunction}
\mathcal{Z}(\beta,\mu_{q},\hat{m}_q) = {\displaystyle \sum_{\substack{\{n_{p},\bar{n}_{p}\} \\ \{k_{\ell},f_{\ell},m_{x} \}  }}}\sigma_{f}\!\!\!\!\!\!{\displaystyle \sum_{\{\rho^{x}_{\pm\mu}\big| \rho^{x}_{\mu} = \rho_{-\mu}^{x+\mu}\}}}\!\!\!\prod_{p}\frac{\tilde{\beta}^{n_{p}+\bar{n}_{p}}}{n_{p}!\bar{n}_{p}!}\prod_{\ell=(x,\mu)}\frac{e^{\mu_{q}\delta_{\mu,0}f_{x,\mu}}}{k_{\ell}!(k_{\ell}+|f_{\ell}|)!}{\displaystyle \prod_{x}}\frac{(2\hat{m}_q)^{m_{x}}}{m_{x}!}T_{x}^{\rho^{x}_{-d},\dots,\rho_{d}^{x}}(\mathcal{D}_{x}),
\end{equation}
\end{widetext}
where the staggered phases $\eta_{\mu}$ are included in the fermionic sign $\sigma_{f}$ whose form will be discussed in the next subsection.
In Eq.~(\ref{DualizedPartitionFunction}) the dependence of the \rhonames{} $\rho^{x}_{\mu}$ on $\{n_{p},\bar{n}_{p},k_{\ell},f_{\ell},m_{x}\}$ is implicit, and the constraints in Eqs.~(\ref{GaugeConstraint}),~(\ref{GrassmanConstraint}) are supposed to be fulfilled. In Fig.~\ref{Fig3} we show the typical structure of an allowed configuration in $d=2$ for $N=3$. \rhonames{} are not shown. Notice that quark fluxes $f_{x,\mu}$ always form closed loops due to the flux conservation law in Eq.~(\ref{GrassmanConstraint}). As opposed to the strong coupling limit, the loops can overlap with dimers and can be intersecting. The system is thus an ensemble of unoriented dimers $k_{\ell}$, monomers $m_{x}$, closed quark fluxes $f_{\ell}$ and plaquettes. The \rhonames{} instead, can be either thought as a mere mathematical tool to automatize the computation of the statistical weights away from strong coupling or as an additional degrees of freedom to be also sampled via Monte Carlo. Before discussing these two possibilities, we want to highlight some features of the partition function Eq.~(\ref{DualizedPartitionFunction}).  \\

\begin{center}
\begin{figure*}
\includegraphics[scale=0.17]{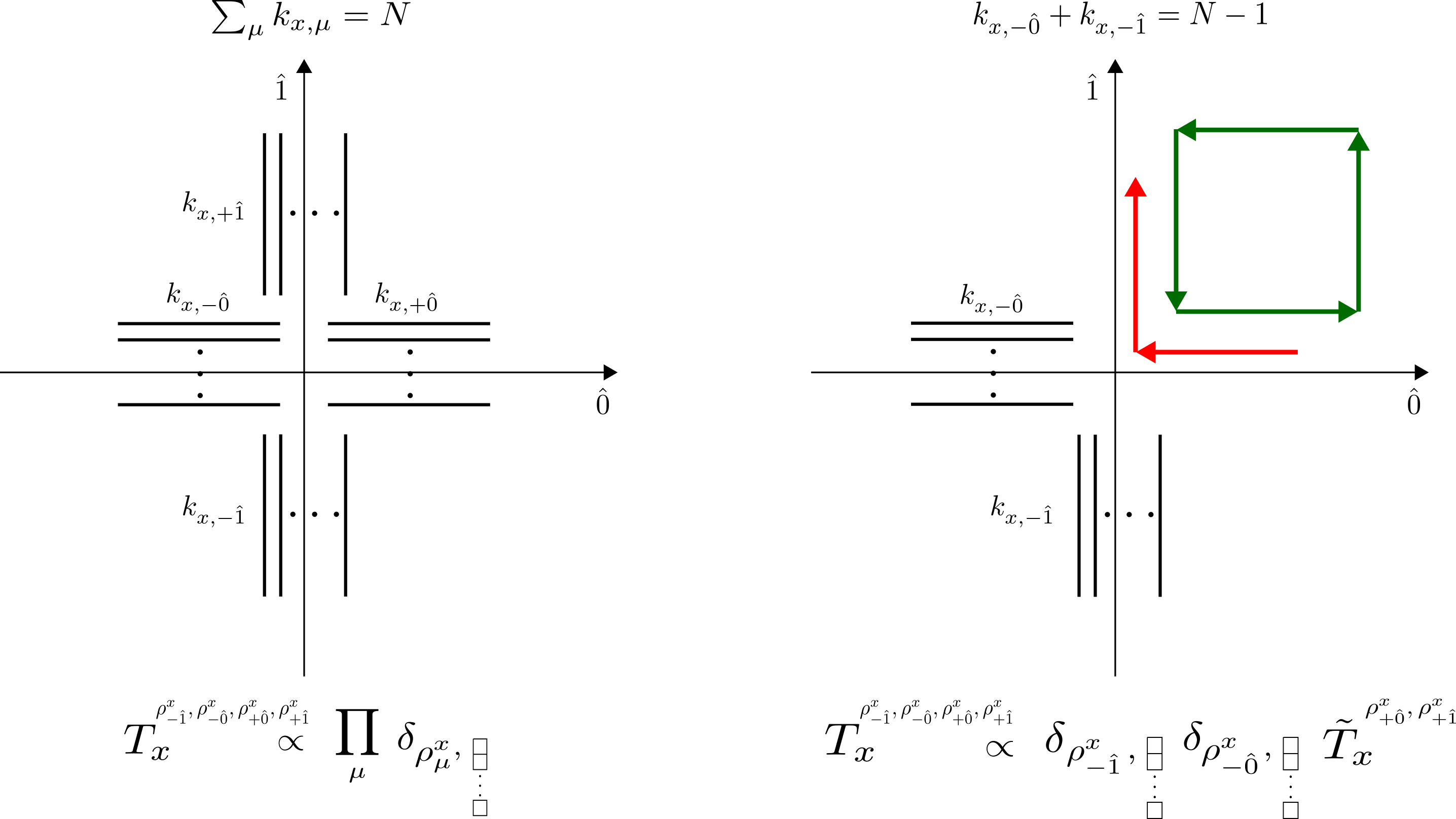}
\caption{\emph{Left}: A typical strong coupling configuration where dimers are attached to a given site. The tensor $T^{\rho}$ is trivial as only one combination of indices (totally antisymmetric irrep on each leg) contributes. \emph{Right}: An $O(\beta)$ correction. The tensor $T^{\rho}$ can be written as external product of a tensor carrying only the \rhonames{} from excited links and delta functions corresponding to the strong coupling legs. }
\label{Fig4}
\end{figure*}
\end{center}

A great simplification occurring, is that the strong coupling contributions always decouple from those corresponding to non-zero $\{n_{p},\bar{n}_{p}\}$. As we 
showed in~\cite{Gagliardi:2017uag}, at strong coupling the tensors $T^{\rho}$ have only one non-zero element. Although for baryon fluxes $(f_{x,\mu}= \pm N)$ this is a trivial statement as there is only one possible \rhoname{} per link, in the case of dimer contributions it is a consequence of the structure of the Decoupling Operators. To show this feature, let us consider the case where only dimers are attached to a given site (Fig.~\ref{Fig4}, left). Contracting the indices of each delta function appearing in the definitions of the corresponding operators Eq.~(\ref{Final-I-Integral}) (for dimer contributions epsilon tensors are absent) with the Grassmann fields, we obtain:
\begin{align}
\label{Sc_derivationI}
\int [d\bar{\chi} d\chi] 
{\displaystyle \prod_{\mu < 0}}
[\bar{\chi}^{k}\chi_{j} \delta_{k_{\pi}}^{\;j}]_{\mu}
{\displaystyle \prod_{\mu>0}} [\bar{\chi}^{i}\chi_{l}
\delta_{i}^{\;l_{\pi}}]_\mu = N!{\displaystyle \prod_{\pm \mu}} \sgn(\pi_{\mu}),
\end{align}
where $\sgn(\pi_{\mu})$ is the parity of the permutation $\pi$ relative to the operator $P$ in direction $\mu$. Hence, the contraction of single deltas decouples, and due to the \textit{Great Orthogonality theorem} (GOT) the only surviving \rhoname{} is the one associated to the totally antisymmetric irrep of the symmetric group:
\begin{align}
\label{Sc_derivationII}
T_{x}^{\rho^{x}_{-d},\dots,\rho^{x}_{d}} &=N!\prod_{\pm\mu}\!\bigg({\displaystyle \sum_{\pi_{\mu}}}\frac{1}{k_{x,\mu}!}\frac{f_{\lambda_{\mu}}}{\sqrt{D_{\lambda_{\mu},N}}}M^{\lambda_{\mu}}_{m,n}(\pi_{\mu})\sgn(\pi_{\mu})\!\!\bigg)\nonumber \\
 &= N!\prod_{\pm\mu}\frac{1}{\sqrt{D_{\lambda_{\mu},N}}}\;\delta_{\lambda_{\mu},\;\tiny\Yvcentermath4\yng(1,1,1)\;\big\rbrace{\;k_{x,\mu}}},\\
D_{\,\,\tiny\Yvcentermath4\yng(1,1,1)\big\rbrace{\;k_{\ell}},N}&=\frac{N(N-1)\cdots (N-k_{\ell}+1)}{k_\ell!},
\end{align}
where in this case $\rho^{x}_{\mu}= (m,n)_{\lambda_{\mu}}$ as there are no epsilon tensors. Given this result, one can obtain the usual contributions from monomers and dimers ($f_{\ell}=0$) to the strong coupling partition function:
\begin{align}
\label{Strong_coupling_part_function}
\mathcal{Z}_{SC}&=\!\!{\displaystyle \sum_{\!\{k_{\ell},m_{x}\}}}\,{\displaystyle \prod_{\ell=(x,\mu)}}\frac{1}{k_{\ell}!^{2}}\,{\displaystyle \prod_{x}}\frac{N!}{m_{x}!}\prod_{\pm\mu}\frac{1}{\sqrt{D_{\lambda_{\mu},N}}}\;\delta_{\lambda_{\mu},\;\tiny\Yvcentermath4\yng(1,1,1)\;\big\rbrace{k_{x,\mu}}}\nonumber \\
&=\!\!{\displaystyle \sum_{\!\{ k_{\ell},m_{x}\}}}\,{\displaystyle \prod_{\ell=(x,\mu)}}\frac{1}{k_{\ell}!^{2}}\,\frac{1}{D_{\,\,\tiny\Yvcentermath4\yng(1,1,1)\;\big\rbrace{\;k_{\ell}},N}}{\displaystyle \prod_{x}}\frac{N!}{m_{x}!} \nonumber \\
&=\!\!{\displaystyle \sum_{\!\{ k_{\ell},m_{x}\}}}\,{\displaystyle \prod_{\ell=(x,\mu)}}\frac{(N-k_{\ell})!}{k_{\ell}!\; N!}{\displaystyle \prod_{x}}\frac{N!}{m_{x}!},
\end{align}\\
where we dropped the dependency on $\hat{m}_q$ and $\mu_{q}$ in the partition function Eq.~(\ref{DualizedPartitionFunction}) at $\beta=0$. The weight for strong coupling baryon loops can be also easily recovered since the corresponding tensors are of size one by construction. 
The decoupling Eq.~(\ref{Sc_derivationII}), also extends to the case where strong coupling dimers combine on a given site with links carrying a non-zero gauge flux. In this case the tensor $T^{\rho}$ can be decomposed as:
\begin{equation}
\label{decomposition}
T_{x}^{\rho^{x}_{-d},\dots,\rho^{x}_{d}} \propto {\displaystyle \prod_{\mu \in s.c.}}\delta_{\lambda_{\mu},\;\tiny\Yvcentermath4\yng(1,1,1)\;\big\rbrace{\;k_{x,\mu}}}\;\tilde{T}_{x}^{\rho^{x}_{exc.}},
\end{equation}
where the proportionality coefficient depends on the external strong coupling dimer legs. An example is provided in Figure~\ref{Fig4} (right), while in Appendix~[\ref{App2}] we re-derive the $\mathcal{O}(\beta)$ partition function. The indices of the tensor in the r.h.s. of Eq.~(\ref{decomposition}) $(\rho^{x}_{exc.})$ correspond to the \rhonames{} of the links attached to excited plaquettes. A similar decomposition holds in presence of an external baryon. As a consequence, the value of $\boldsymbol{\mathcal{G}}$ can be written as:
\begin{equation}
\label{Bubble_decomposition}
\mathcal{G} = \mathcal{G}_{s.c.}\;{\displaystyle \prod_{\text{bubbles i}}}\mathcal{G}_{\mathcal{B}_{i}},
\end{equation}
where a bubble $\mathcal{B}_{i}$ is any \textit{plaquette-connected} region and two bubbles are disconnected if they do not share an excited link (i.e. a link attached to an excited plaquette).\\
Therefore, to evaluate the total weight of a configuration, it is sufficient to use the more involved structure based on the tensor network contraction on the sublattice where the plaquette occupation numbers are non-zero, exploiting the factorization of the tensor network for disconnected plaquette contributions. The strong coupling part can be evaluated using the standard combinatorial formulae (e.g. Eq.~[\ref{Strong_coupling_part_function}]). This is particularly useful since at small values of $\beta$ the bubbles $\mathcal{B}_{i}$ extend over few lattice spacings and the non-local effects from the tensor network are manageable.\\
\subsection{Complexity and sampling strategies}
\label{Complexity}
\begin{center}
\begin{figure*}
\includegraphics[scale=0.154]{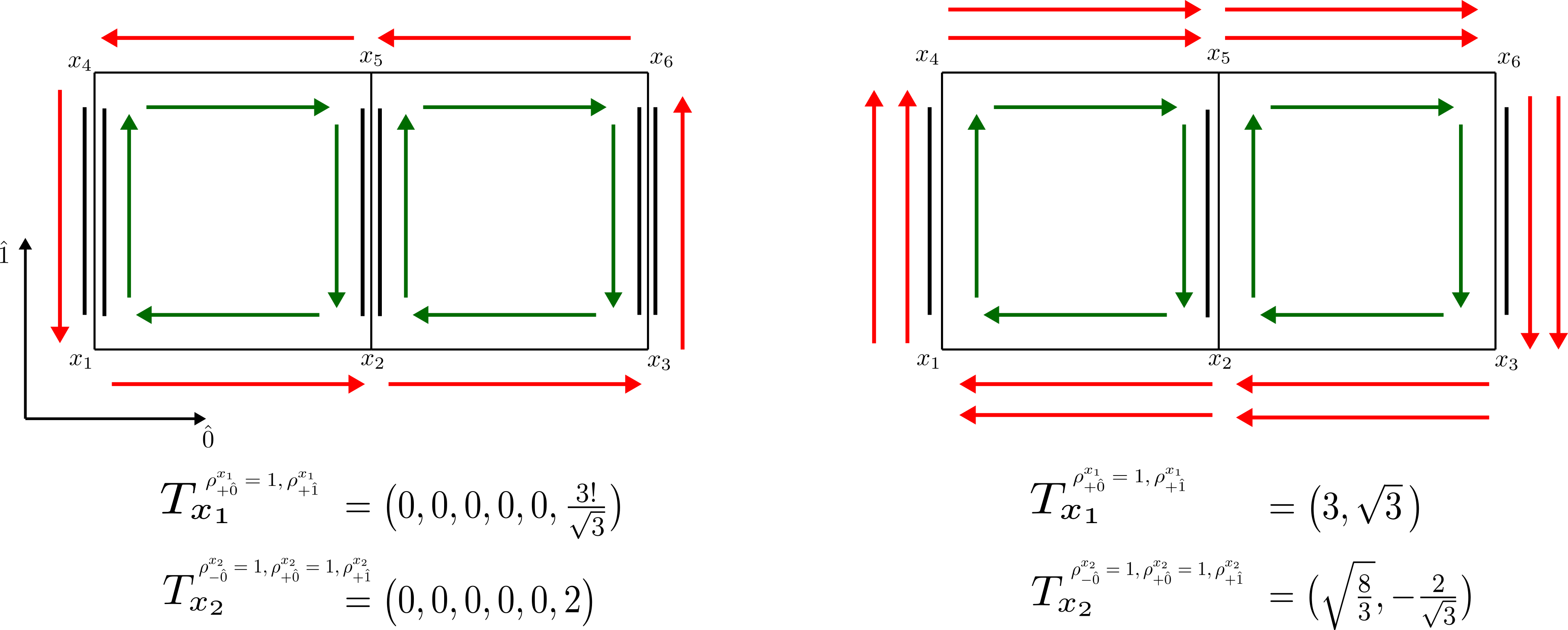}
\caption{Two $\SU(3)$ bubble contributions at $\mathcal{O}(\beta^{2})$ with $\left(n_{p},\bar{n}_{p}\right) = (0,1)$ on the two excited plaquettes. At this order, computing the weight corresponding to the bubbles is easy as the tensor network within the bubble is only made up of small vectors. Some external legs are trivial as there is only one possible \rhoname{} $(\rho^{x}_{\mu}=1)$. Oftentimes the tensors $T_{x}$ are very sparse as a consequence of the Great Orthogonality Theorem (see also Appendix.~\ref{App2}). In the figures we only show the tensors associated to $x_{1}$ and $x_{2}$. The remaining tensors are given by: $T_{x_{3}}= T_{x_{4}}=T_{x_{6}}=T_{x_{1}}$ and $T_{x_{5}}= T_{x_{2}}$. }
\label{Fig7}
\end{figure*}
\end{center}

We now want to comment on the complexity of the dual partition function Eq.~(\ref{DualizedPartitionFunction}). Given the background $\{n_{p},\bar{n}_{p}, k_{\ell}, f_{\ell},m_{x}\}$, the weight of the configuration is obtained by contracting the tensor network $T_{x}^{\rho}$.  Two different strategies can be used to sample the partition function: one can either exploit the "bubble decomposition" in Eq.~(\ref{Bubble_decomposition}) to simplify the numerical cost of contracting the network, or consider the \rhonames{} as an additional degrees of freedom to be evaluated stochastically.  In the first case a relevant question is whether the decomposition Eq.~(\ref{Final-I-Integral}) is optimal, meaning that the number of Decoupling Operators $n_{\rho}$ in Eq.~(\ref{numberOfOperators}) is the smallest possible. The machine time required to contract the network depends almost completely on the size of the external legs of the tensors. In the $\U(N)$ case we already know that the answer is positive as it can be shown that the operators $P^{\rho}$ are mutually orthogonal, hence independent. It is therefore not possible to perform a reparameterization of the $\mathcal{I}$-integral that results in a decomposition of the type Eq.~(\ref{Final-I-Integral}) with a smaller number of terms within the sum. For $\SU(N)$ the situation is not completely clear as we could not prove that the Decoupling Operators corresponding to the $\SU(N)$ contributions (non-zero $q$) are independent. The question whether the complexity can be reduced using a different parameterization is thus still open. In any case, the lower bound on the number of \rhonames{} provided by the $\U(N)$ result already tells that to a certain degree, the complexity is unavoidable. This number grows as a factorial as the (anti-) plaquette occupation numbers increase and contracting the resulting tensors along the excited plaquettes becomes in general too expensive in $d>2$. Even though this description can be used as a starting point for future theoretical development, as it stands, the bubble decomposition and the corresponding tensor network cannot be used for exact calculations in full QCD. Nevertheless, the dual form of the partition function together with the decomposition Eq.~(\ref{Bubble_decomposition}) can be used to study lattice QCD perturbatively in $\beta$, by truncating the expansion of the plaquette action. We remind that this has been done so far, using worldline formulations, only for the leading $\mathcal{O}(\beta)$ corrections~\cite{deForcrand2014}. Truncating at $\mathcal{O}(\beta^{n})$ means that the allowed configurations are only those corresponding to bubble contributions of at most $\mathcal{O}(\beta^{n})$. Making use of this definition, the truncation corresponds to a free energy which is exact up to the same order. For instance, at order $\mathcal{O}(\beta^{2})$, the largest allowed bubble contributions are $2\times 1$ rectangles with an elementary (anti-) plaquette excitation ($\bar{n}_{p}$) $n_{p}=1$ as sketched in Fig.~\ref{Fig7} In the $\SU(3)$ case, four of the six tensors $T_{x}^{\rho}$ making up the bubble, are matrices of sizes at most $6\times 1 $, while the other two are rank 3 tensors of sizes at most $6\times 1\times 1$. Contracting the reduced tensor network within the bubbles is straightforward and can be done on the fly during Monte Carlo evolution without any overhead. Higher order contributions ($n=3,4,5,...$) can be also easily evaluated in $4d$. One possible strategy is to compute and store beforehand all the tensors $T^{\rho}$ that are compatible with the constraint and the truncation order. This step needs to be performed one time only, as the tensor network does not depend on the simulation parameters. For instance, the computation of all the tensors needed to address the $4d$ $N^{3}LO$ correction to strong coupling QCD took $\approx 10^{2}s$ on a single CPU, with the largest tensor having only $\mathcal{O}(10)$ non-zero elements. The tensors are then loaded and used to compute the value of $\mathcal{G}_{B_{i}}$ when the bubble $B_{i}$ needs to be updated. We are currently designing an ergodic algorithm capable to sample the bubble contributions which will be illustrated in a forthcoming publication where the higher order $\beta$ corrections to the strong coupling phase diagram will be addressed. 

The second possibility is to consider the DOIs as an additional degrees of freedom along with $\{n_{p},\bar{n}_{p},k_{\ell},m_{x}\}$. The complexity of the tensor network can be thus overcome by importance sampling. In this case, a given configuration is determined by selecting one tensor element for each lattice site. When doing so, the weight of such configurations is local and an additional Metropolis acceptance test can be easily introduced to make sure that the system explores the DOIs configuration space during Monte Carlo. For instance, when a bond, an elementary plaquette or a cube containing $6$ plaquettes is updated, we can propose a quasi-local update by randomly choosing  new \rhonames{} on the bonds involved. The feasibility of this approach depends on the minus signs induced by splitting the former configurations in terms of $\{n_{p},\bar{n}_{p},k_{\ell},m_{x}\}$ into sub-configurations where one selects a single tensor element out of the full tensor $T^{\rho}_{x}$. In fact, the tensor elements are not positive defined and it could happen that without contracting the network an additional source of minus signs is plugged into the system. As mentioned in Sec.~\ref{PartitionFunction}, the main obstacle to the use of the permutation basis was in fact the severe sign problem induced by the Weingarten functions $\tWg$. Using instead the DOIs, the induced sign problem can be much milder. In Sec.~\ref{IV} we will provide preliminary evidences to this statement based on an exact enumeration of the partition function.

\subsection{Sign Problem}
Having discussed the partition function, we now turn to the computation of $\sigma_{f}$ in the dual representation. In general, the fermionic sign of a configuration is determined by the staggered phases, the anti-periodic boundary condition for fermion fields and by the so-called geometric sign. The latter stems from the fact that, starting from Eqs.~(\ref{SCE}),~(\ref{G}), one has to reorder the Grassmann variables contained in the matrices $\mathcal{M}_{\ell},\mathcal{M}^{\dag}_{\ell}$ before performing the Grassmann integration at each site. At strong coupling, only baryon loops ($f_{\ell}=\pm N$) can induce a negative sign and the geometric sign is known in closed form. It combines with the staggered phases and the winding number to produce
\begin{align}
\label{sign}
\sigma_{f}(\mathcal{C}) = \left[{\displaystyle \prod_{\ell\in \mathcal{C}}}\eta_{\mu}(x)\right]\,\left(-1\right)^{N_{\ell}(\mathcal{C})+N_{\_}(\mathcal{C})+ \omega(\mathcal{C})}
\end{align} 
for $\SU(2N+1)$, whereas $\sigma_{f}=+1$ for $\SU(2N)$. In Eq.~(\ref{sign}), $\mathcal{C}$ is the set of links traversed by baryons, $N_{\ell}(\mathcal{C})$ the number of baryon loops, $N_{\_}(\mathcal{C})$ the number of baryon loop segments in negative directions and $\omega(\mathcal{C})$ the total winding number in temporal direction. At strong coupling, the baryon-loop induced sign problem is very mild and the finite density phase diagram can be mapped out using sign reweighting~\cite{Forcrand2010,Unger2011}. \\

At finite $\beta$ the structure of the geometric sign gets more complicated as the allowed quark fluxes can also be intersecting and the equality in Eq.~(\ref{sign}) does no longer hold true. Specializing to $\SU(3)$, a fermionic minus sign is only induced by single and triple quark fluxes while for dimers and di-quarks $\sigma_{f}=+1$\footnote{For $\SU(2)$ only single quark fluxes can produce a negative $\sigma_{f}$.}. To compute the geometric sign for intersecting loops, as closed formulae are apparently lacking, we 
explicitly count how many times the Grassmann variables corresponding to odd fluxes need to be commuted to bring them in canonical ordering at each lattice site. Formally, $\sigma_{f}$ can be written as
\begin{equation}
\label{FermionicSign}
\sigma_{f}(\mathcal{C}_{1},\mathcal{C}_{3}) = \bigg[{\displaystyle \prod_{\,\ell \in \mathcal{C}_{1}\cup\mathcal{C}_{3}}}\eta_{\mu}(x)\bigg](-1)^{\omega(\mathcal{C}_{1},\mathcal{C}_{3})}\,\sigma_{G}(\mathcal{C}_{1},\mathcal{C}_{3}),
\end{equation} 
where $\mathcal{C}_{1}$ and $\mathcal{C}_{3}$ are respectively the set of links traversed by single and triple quark fluxes and the winding number $\omega(\mathcal{C}_{1},\mathcal{C}_{3})$ is given by
\begin{equation}
\omega(\mathcal{C}_{1},\mathcal{C}_{3}) = \sum_{\vec{x}}f_{(\vec{x},N_{\tau}), \hat{0}},
\end{equation}
where $N_{\tau}$ is the temporal extent of the lattice. In Eq.~(\ref{FermionicSign}), $\sigma_{G}$ is the global geometric sign and in general cannot be factorized as a product of two terms depending separately on $\mathcal{C}_{1}$ and $\mathcal{C}_{3}$. It is computed after contracting the tensor $T_{x}^{\rho}$ at fixed background $\{n_{p},\bar{n}_{p},d_{\ell},f_{\ell},m_{x}\}$, and cannot be cast into a product of local minus signs that can be absorbed with a redefinition of the tensors $T_{x}$. \\

As we already mentioned, another potential source of negative signs, which does not depend on the fermion fields, is caused by the lack of positivity of the tensor elements $T_{x}^{\rho}$. This issue is relevant when considering the \rhonames{} as an additional degrees of freedom. Strong oscillations of the sign within the tensor network can in fact hinder the application of importance sampling. Although, this question can be only answered on the basis of Monte Carlo simulations via sign reweighting, in Sec.~\ref{IV} we will show preliminary results on the interplay between the fermionic and the tensor network induced sign problem, obtained from exact enumeration of the partition function on small volumes.
\subsection{Observables}
As both the fermion field and the gauge links have been integrated out, the observables in the dual representation take a different form. The ones defined as derivatives of $\log\mathcal{Z}$ with respect to external parameters, can be obtained taking derivatives in Eq.~(\ref{DualizedPartitionFunction}). For instance, the chiral condensate $\langle \bar{\psi}\psi\rangle$, baryon number $n_{B}$ and average plaquette $\langle P\rangle$ are given by:
\begin{align}
\label{Observables}
\langle \bar{\psi}\psi\rangle &= \frac{1}{V}\frac{\partial \log\mathcal{Z}}{\partial \hat{m}_q} = \frac{\langle m_{x}\rangle}{V \hat{m}_{q}}, \nonumber \\
\langle n_{B} \rangle &= \frac{1}{V}\frac{\partial \log\mathcal{Z}}{\partial \mu_{q}}= \frac{\langle f_{x,\hat{0}}\rangle}{V}, \nonumber \\
\langle P \rangle &= \frac{1}{N\, V}\frac{\partial \log \mathcal{Z}}{\partial \beta} = \frac{\langle n_{p}+\bar{n}_{p}\rangle}{2\beta V},
\end{align}
and higher order derivatives (i.e. susceptibilities) can be obtained in a similar fashion, evaluating the various cumulants of $m_{x},f_{x,\hat{0}}$ and $n_{p}+\bar{n}_{p}$.
The definition of non-derivative observables, such as the Polyakov loop, is less trivial in the dual representation as the gauge fields have been already integrated out. Formally, the Polyakov loop can be written as a ratio of partition functions
\begin{align}
\langle L \rangle = \frac{\mathcal{Z}_L}{\mathcal{Z}},
\end{align}
where $\mathcal{Z}_L$ is the partition function with a Polaykov loop insertion and $\mathcal{Z}$ is given by Eq.~(\ref{DualizedPartitionFunction}). $\mathcal{Z}_{L}$ admits a dual representation similar to Eq.~(\ref{DualizedPartitionFunction}) with modified tensors $T_{x}^{\rho}$ at the sites $x$ crossed by the Polyakov loop. Here, we will not discuss its specific form. The strategy to  sample the Polyakov loop will be addressed in a following paper containing the numerical results from Monte Carlo simulations.\\

To perform finite temperature calculations at non-zero $\beta$, we can either vary the temporal lattice extent $N_{\tau}$ at fixed lattice spacing $a$ according to
\begin{equation}
aT = \frac{1}{N_{\tau}},
\end{equation}
or perform simulations on anisotropic lattices. The first strategy works well for $\frac{\beta}{2N}$ close to one, where one can meaningfully fix the scale and determine the relations $\beta(a), \hat{m}_q(a)$ imposing a physical constraint on the low-energy mesonic spectrum. Instead, at small enough $\beta$ (and especially at strong coupling), the scale cannot be fixed as the lattice is too coarse. In this case the temperature is changed inducing a physical anisotropy $\xi= \frac{a_{s}}{a_{t}}$ by using two different $\beta$ couplings for spatial ($\beta_{s}$) and temporal ($\beta_{t}$) plaquettes and introducing a fermionic bare anisotropy $\gamma$ that favours hoppings in temporal direction. Implementing this modification in the partition function Eq.~(\ref{DualizedPartitionFunction}) is straightforward. The modifications can be summarized in:
\begin{align}
\frac{e^{\mu_{q}\delta_{\mu,0}f_{x,\mu}}}{k_{\ell}!(k_{\ell}+|f_{\ell}|)!}\quad &\rightarrow \quad\frac{e^{\mu_{q}\delta_{\mu,0}f_{x,\mu}}}{k_{\ell}!(k_{\ell}+|f_{\ell}|)!}\gamma^{\delta_{\mu,0}\left(|f_{x,\mu}|+2k_{x,\mu}\right)}, \nonumber \\
\beta^{n_{p}+\bar{n}_{p}}\quad &\rightarrow \quad\beta_{s}^{n_{p_{s}}+\bar{n}_{p_{s}}}\beta_{t}^{n_{p_{t}}+\bar{n}_{p_{t}}},
\end{align}
and ($\bar{n}_{p_{s/t}}$) $n_{p_{s/t}}$ are the (anti-) plaquette occupation numbers for spatial and temporal plaquettes.
The relation between the bare parameters $\beta_{s}, \beta_{t}, \gamma$ and the physical anisotropy $\xi$ has to be determined non-perturbatively via the so called anisotropy calibration procedure (see~\cite{deForcrand2017} and references therein). This has been done so far in the strong coupling limit at zero~\cite{deForcrand2017} and non-zero~\cite{Bollweg2018} quark mass $\hat{m}_q$. The extension to strong coupling QCD including $\mathcal{O}(\beta)$ is in preparation. In this paper, we will be only interested in the evolution of the observables as a function of $\beta, \hat{m}_q, \mu_{q}$, hence in comparing the dual observables with the HMC results we will set $\gamma=1$ and $\beta_{s}=\beta_{t}=\beta$. 

\section{Crosschecks from Exact Enumeration/HMC for $N=2$, $N=3$ ($\U(N)$, and $\SU(N)$)}
\label{IV}
\onecolumngrid
\begin{figure*}[t]
\centerline{
\includegraphics[page=2,width=0.27\textwidth]{schwinger_02x02_U2_J1.pdf}\hspace{-5mm}
 \includegraphics[page=1,width=0.27\textwidth]{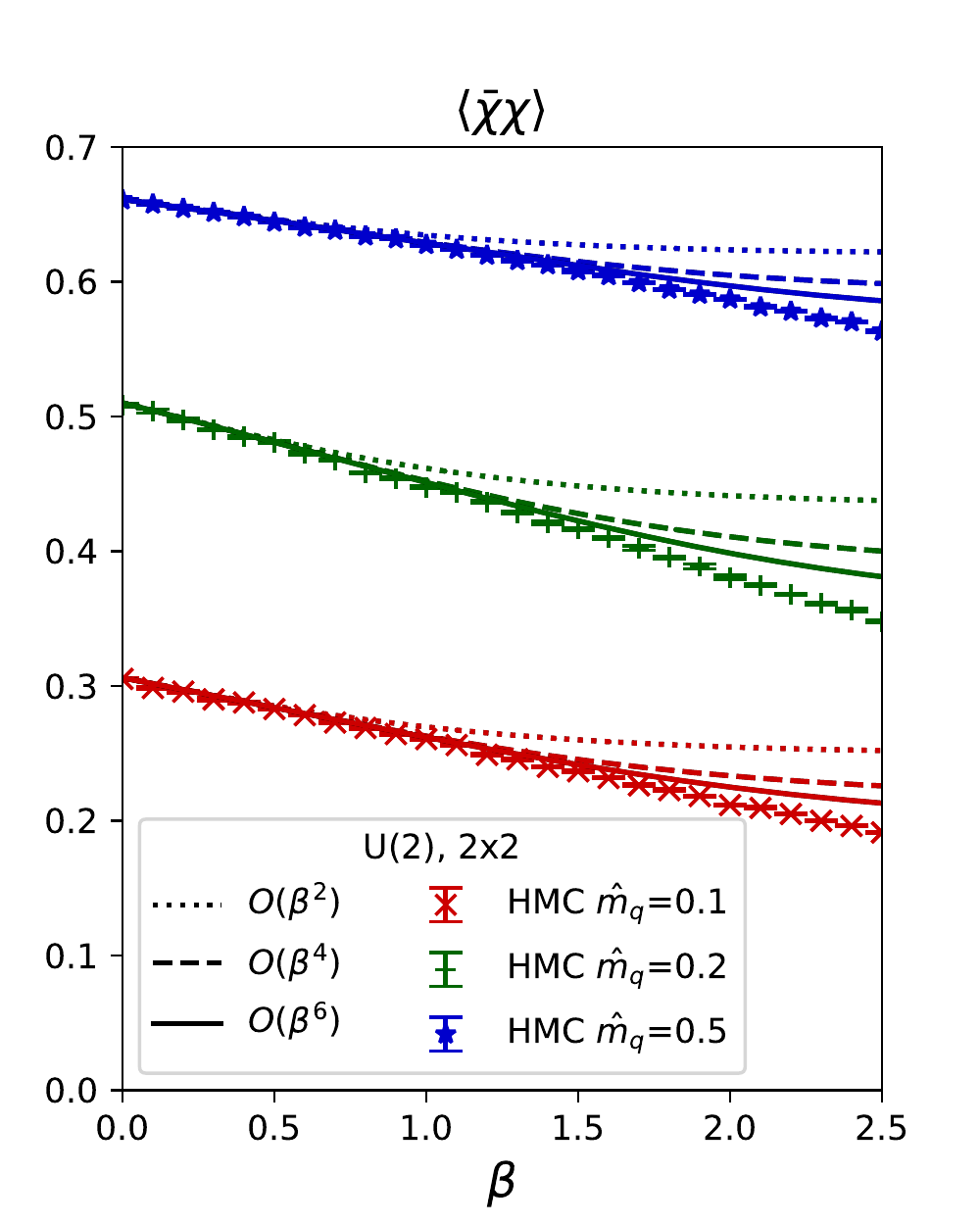}
 \includegraphics[page=2,width=0.27\textwidth]{schwinger_04x04_U2_J1.pdf}\hspace{-5mm}
 \includegraphics[page=1,width=0.27\textwidth]{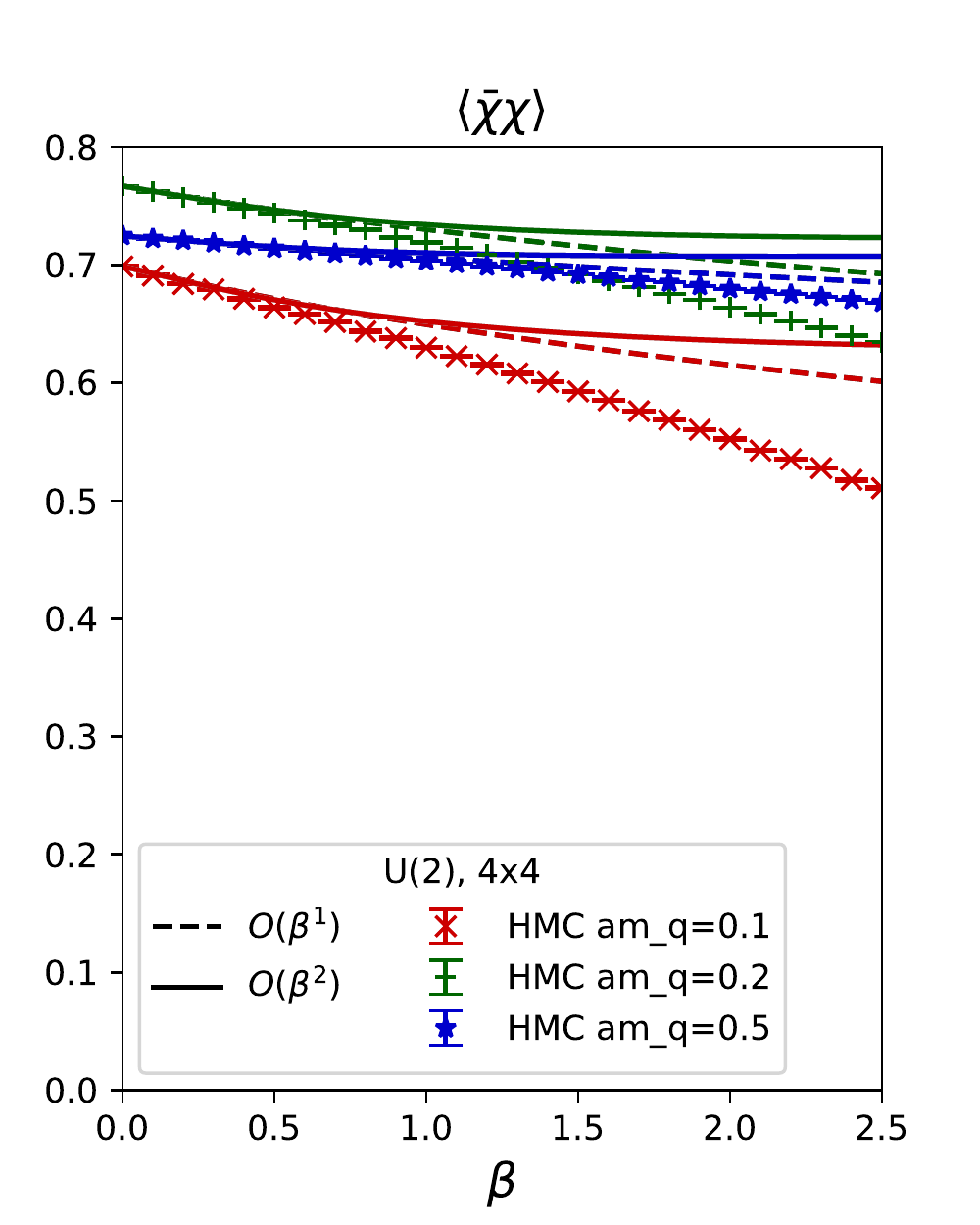}
 }
\centerline{
 \includegraphics[page=2,width=0.27\textwidth]{schwinger_02x02_U3_J1.pdf}\hspace{-5mm}
 \includegraphics[page=1,width=0.27\textwidth]{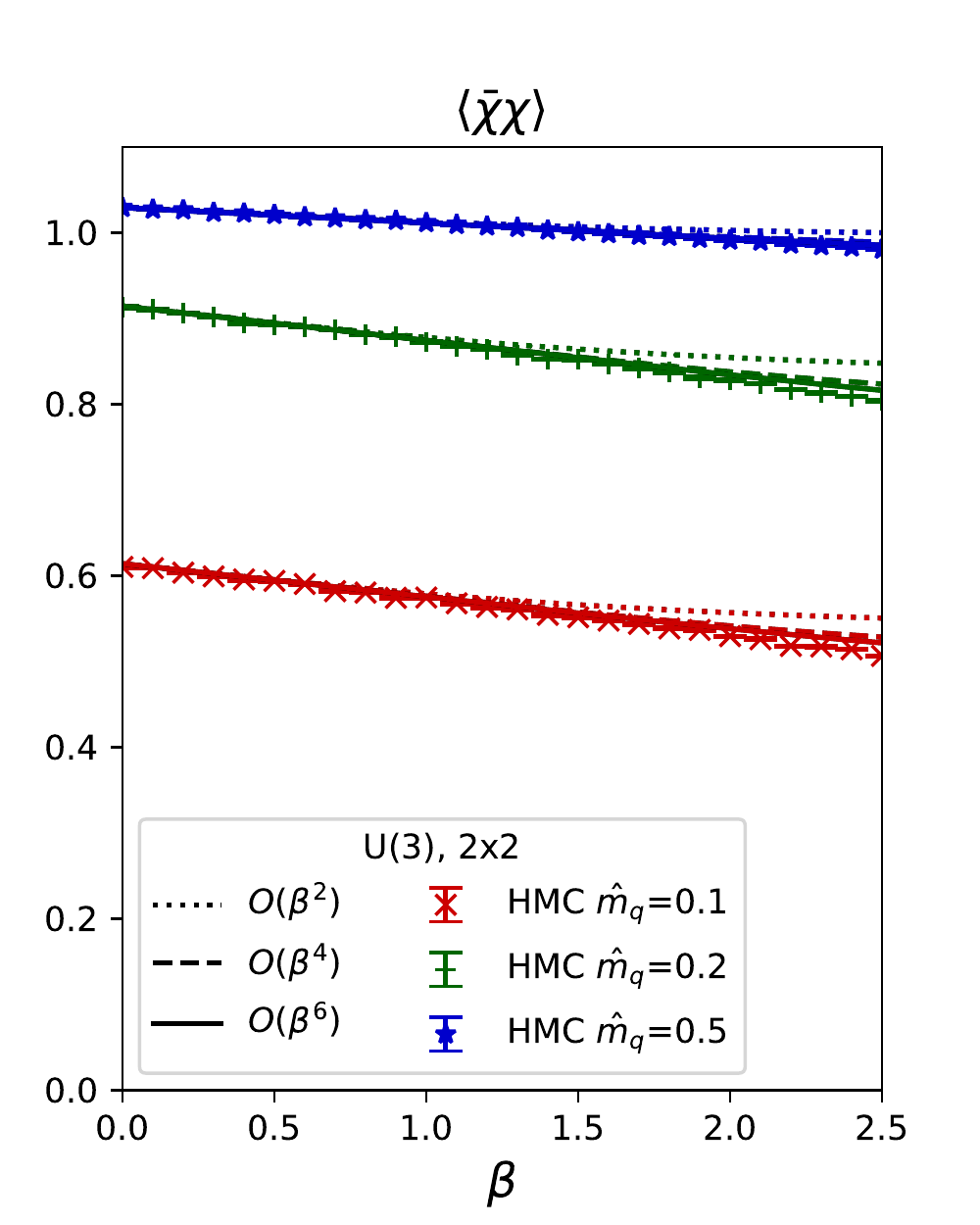}
 \includegraphics[page=2,width=0.27\textwidth]{schwinger_04x04_U3_J1.pdf}\hspace{-5mm}
 \includegraphics[page=1,width=0.27\textwidth]{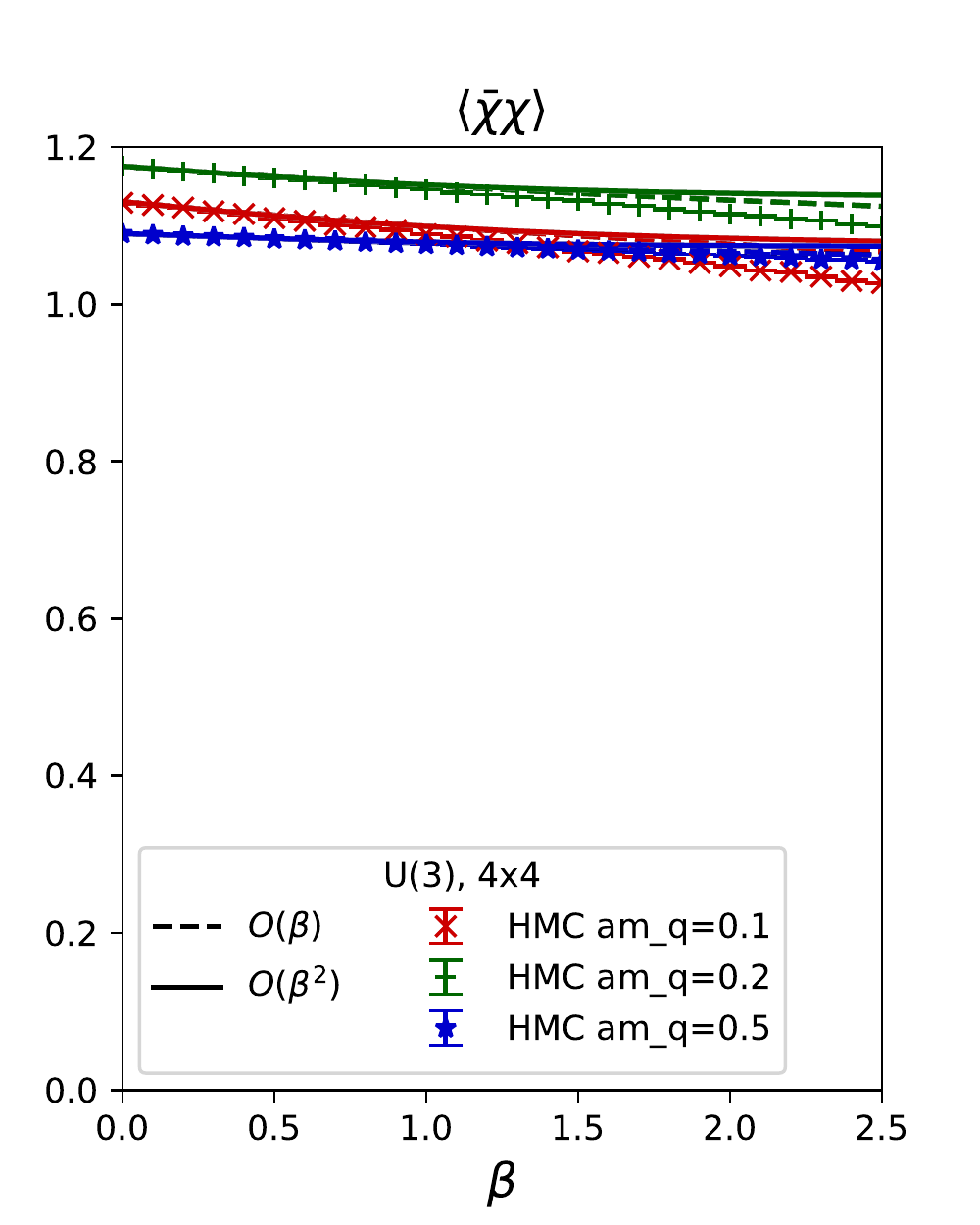}
 }
\caption{Comparison between exact enumeration and HMC simulations for $\U(2)$ \emph{(upper plots)} and $\U(3)$ \emph{(lower plots)}. For both gauge groups the average plaquette $\langle P\rangle$ and the chiral condensate $\langle \bar{\chi}\chi\rangle$ are shown on a $2\times 2$ \emph{(left)} and $4\times 4$ \emph{(right)} lattice. Data points with different symbols correspond to different quark masses while solid, dashed and dotted lines correspond to different truncation orders $\mathcal{O}(\beta^{n})$.   }
\label{U2_U3}
\end{figure*}
\twocolumngrid
In a finite volume, the partition function Eq.~(\ref{DualizedPartitionFunction}) truncated at a given order $\mathcal{O}\left(\beta^{n}\right)$, is always a finite polynomial $\mathcal{P}\left(\beta,\hat{m}_q,z_{q}\right)$ in $\beta$, quark mass $\hat{m}_q$ and fugacity $z_{q}=\exp{\frac{\mu_{q}}{T}}$. To check the correctness of the dual formulation and the computation of the weights, we performed the exact computation of $\mathcal{P}$ in small two-dimensional volumes, comparing the result from the exact enumeration of $\mathcal{Z}$ with the outcome of standard lattice QCD simulations at zero chemical potential $\mu_{q}$. We considered as gauge group both $\U(N)$ and $\SU(N)$ for $N=2,3$, and obtained the full polynomial $\mathcal{P}$ on $2\times 2$ and $4\times 4$ volumes for various $n\leq 6$, employing periodic boundary conditions in all directions. To enumerate the coefficients of the polynomial, we first pre-computed all the tensors $T_{x}(\mathcal{D}_{x})$ compatible with Gauge and Grassmann constraints and with the truncation order. We then generated all possible combinations of $\{n_{p},\bar{n}_{p},d_{\ell},f_{\ell},m_{x}\}$ and contracted the corresponding tensor network to determine its contribution to $\mathcal{P}$. The result of this contraction was then multiplied by the fermionic sign $\sigma_{f}$. In Tab.~\ref{Tab1} we show the total number of configurations as a function of the truncation order. At fixed $\mathcal{O}(\beta^{n})$, this number grows very large as a function of the number of dimensions, hence we could not perform the exact enumeration in $d>2$. 
Nevertheless, as our dual formulation does not present any fundamental difference when applied to higher dimensions, we believe that this crosscheck gives some hints about its validity in $d=3,4$. In Figs.~\ref{U2_U3}-~\ref{SU2_SU3}, we show the results of this comparison for the average plaquette $\langle P\rangle$ and for the chiral condensate $\langle \bar{\chi}\chi\rangle$, respectively for $\U(2)$, $\U(3)$ and $\SU(2)$, $\SU(3)$. They were analytically determined from $\mathcal{P}\left(\beta,\hat{m}_q,z_{q}\right)$ by
\begin{align}
\langle P\rangle &= \frac{1}{NL^{2}}\partial_{\beta}\mathcal{P}\left(\beta,\hat{m}_q,0\right)/\mathcal{P}\left(\beta,\hat{m}_q,0\right), \nonumber \\
\langle\bar{\chi}\chi\rangle &= \frac{1}{L^{2}}\partial_{\hat{m}_q}\mathcal{P}\left(\beta,\hat{m}_q,0\right)/\mathcal{P}\left(\beta,\hat{m}_q,0\right),
\end{align}
for $L=2,4$ and $N=2,3$. A clear result, emerging from Figs.~\ref{U2_U3}~-\ref{SU2_SU3}, is that the strong coupling branch is well described by the polynomials $\mathcal{P}$ for all quark masses, and that the agreement with the HMC results indeed extends to larger and larger $\beta$ as the truncation order is increased. Notice that at any fixed order $\mathcal{O}(\beta^{n})$, the continuum limit $\beta\to\infty$ of the average plaquette is always zero, as it can be seen from its definition in terms of the dual degrees of freedom Eq.~(\ref{Observables}). The value $\beta_{max}$ that corresponds to a maximum of the average plaquette can be used as a strong upper bound for the validity of the expansion at $\mathcal{O}(\beta^{n})$.\\

\onecolumngrid
\begin{figure*}[t]
\centerline{
 \includegraphics[page=2,width=0.27\textwidth]{schwinger_02x02_SU2_J1.pdf}\hspace{-5mm}
 \includegraphics[page=1,width=0.27\textwidth]{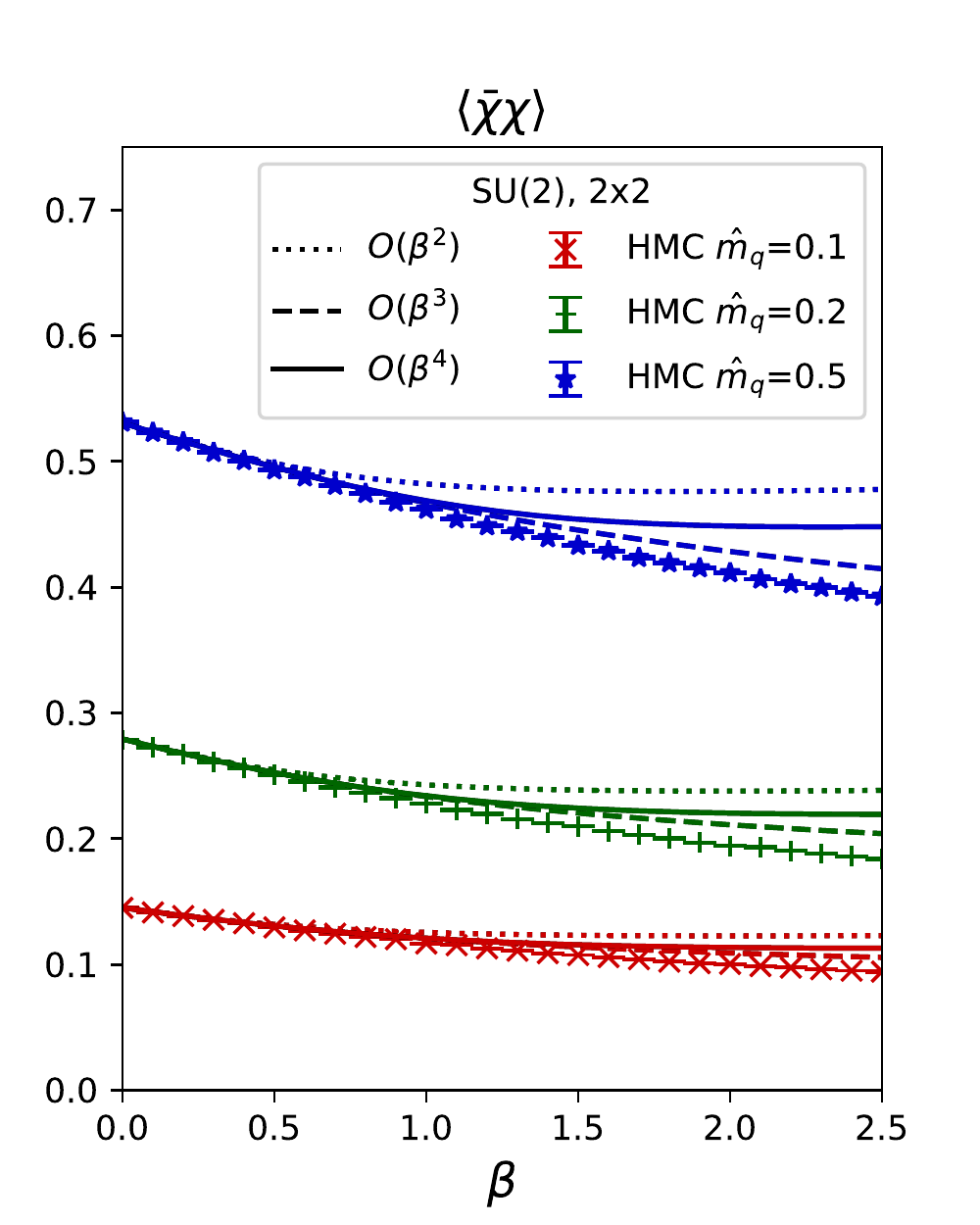}
 \includegraphics[page=2,width=0.27\textwidth]{schwinger_02x02_SU3_J1.pdf}\hspace{-5mm}
 \includegraphics[page=1,width=0.27\textwidth]{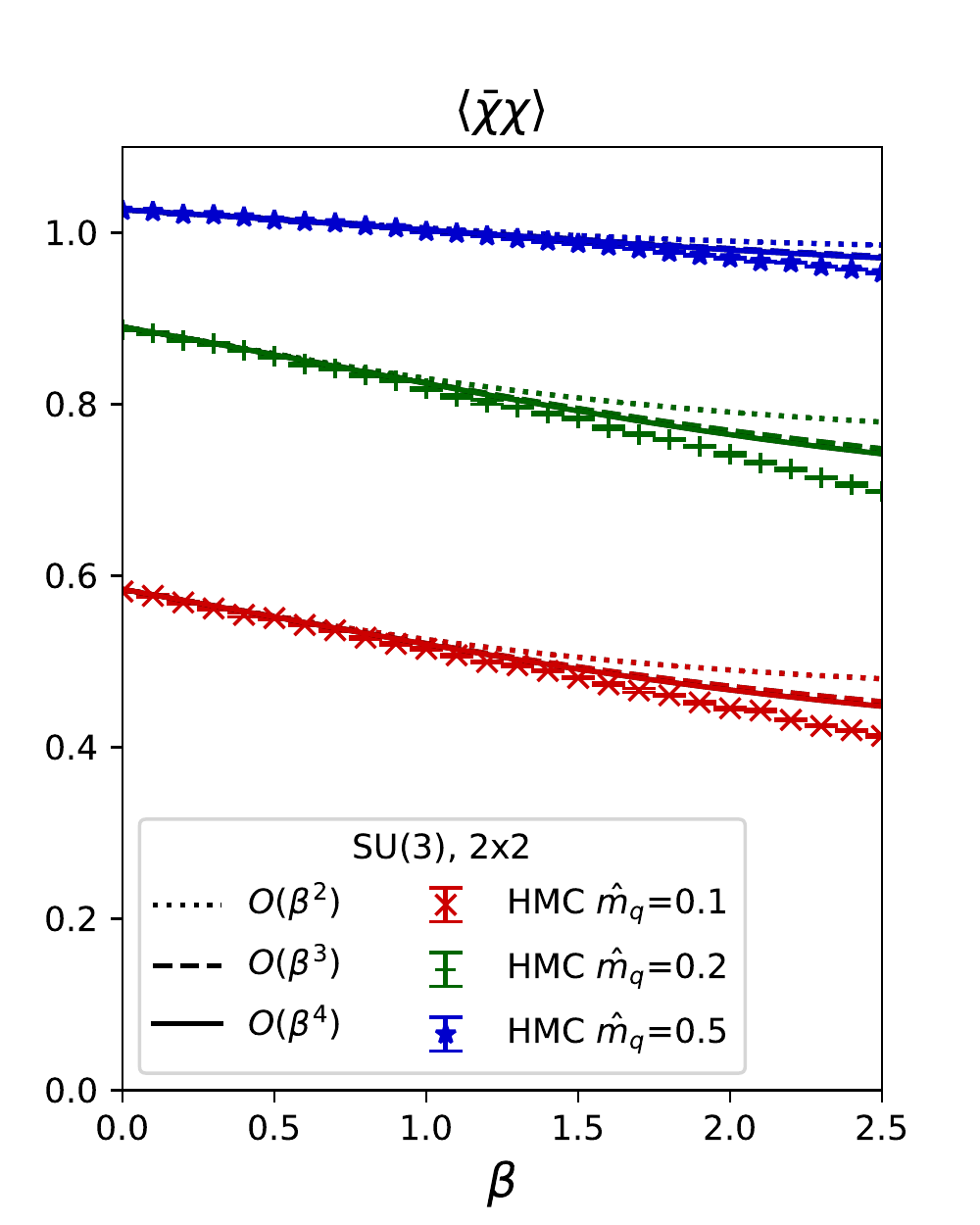}
 }
\caption{Similar comparison between exact enumeration and HMC on a $2\times 2$ lattice as in Fig.~\ref{U2_U3}, for $\SU(2)$ \emph{(left plots)} and $\SU(3)$ \emph{(right plots)}.}
\label{SU2_SU3}
\end{figure*}
\twocolumngrid

\setlength{\tabcolsep}{4pt}
\begin{table}[b!]
\vspace{5mm}
\begin{center}
\begin{tabular}{|l|rrrrr|}
\hline
 & U(1) & U(2) & U(3) & SU(2) & SU(3) \\
\hline
$\mathcal{O}(\beta^0)$ &  17 &   135   &    695 &     223 &  815\\
$\mathcal{O}(\beta^1)$ &  25 &   271   &   1\,775 &     863 &  2\,495\\
$\mathcal{O}(\beta^2)$ & 101 &  1\,839 &  12\,163 &   14\,471 &  25\,259\\
$\mathcal{O}(\beta^3)$ & 141 &  4\,119 &  36\,027 &  152\,551 &  337\,503\\
$\mathcal{O}(\beta^4)$ & 373 & 32\,107 & 436\,415 & 4\,895\,849 & 4\,703\,047 \\
$\mathcal{O}(\beta^5)$ & 497 & 80\,319 & 1\,640\,829 & 106\,758\,281 & 182\,863\,979 \\
\hline
\end{tabular} 
\caption{Number of distinct configurations on a $2\times2$ lattice, 
for various gauge groups and truncations of $\mathcal{O}(\beta^n)$. All these configurations are taken into account when computing the partition function and its derivatives, as shown in Figs.~\ref{U2_U3},~\ref{SU2_SU3}. The complexity of enumeration rises drastically with $n$, but can be overcome by importance sampling.
\label{Tab1}
}
\end{center}
%\end{table}
%\begin{table}
\begin{tabular}{|l|rrrrr|}
\hline
 & U(1) & U(2) & U(3) & SU(2) & SU(3) \\
\hline
$\mathcal{O}(\beta^0)$ &   5 &   15 &    35 &     27 &  47\\
$\mathcal{O}(\beta^1)$ &  13 &   55 &   155 &     155 &  255\\
$\mathcal{O}(\beta^2)$ &  41 &  215 &  655 &   1139 &  1499\\
$\mathcal{O}(\beta^3)$ &  81 &  639 &  2279 &  6995 &  8939\\
$\mathcal{O}(\beta^4)$ & 173 & 2079 & 8687 & 48957 & 52571 \\
$\mathcal{O}(\beta^5)$ & 293 & 6007 & 31617 & 338109 & 360525 \\
\hline
\end{tabular} 
\caption{Number of distinct non-zero tensor elements 
for various gauge groups and truncations of $\mathcal{O}(\beta^n)$ in two dimensions. These numbers correspond to the total number of vertices that would enter in a corresponding vertex model. 
\label{Tab2}
}
\end{table}

\vphantom{abcd}

\vphantom{abcd}

\begin{figure*}
\centerline{
 \includegraphics[page=3,width=0.27\textwidth]{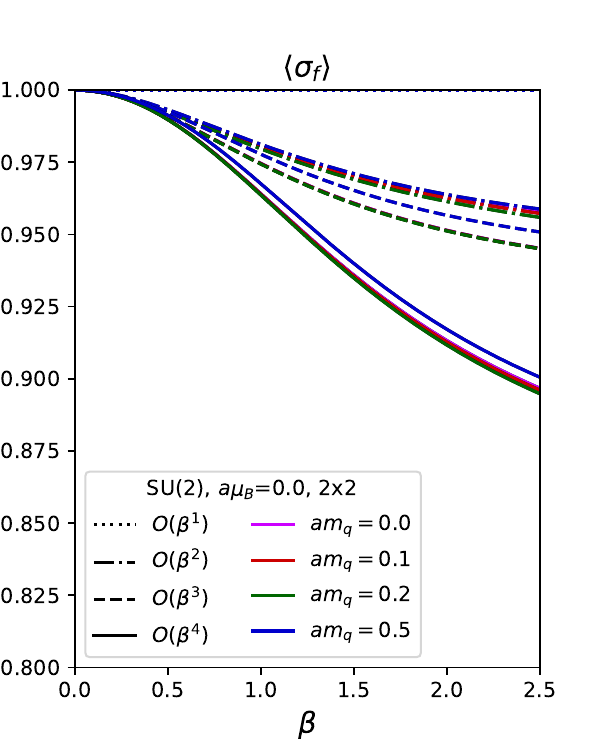}\hspace{-5mm}
 \includegraphics[page=3,width=0.27\textwidth]{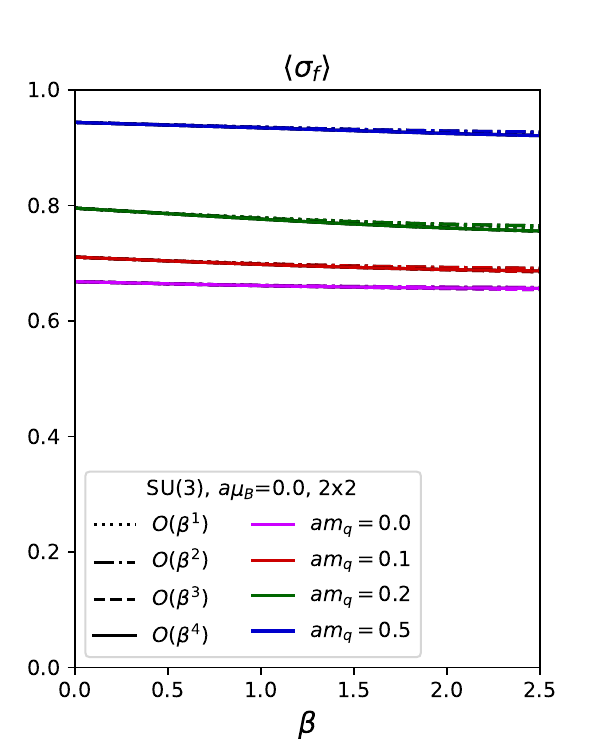}
 \includegraphics[page=3,width=0.27\textwidth]{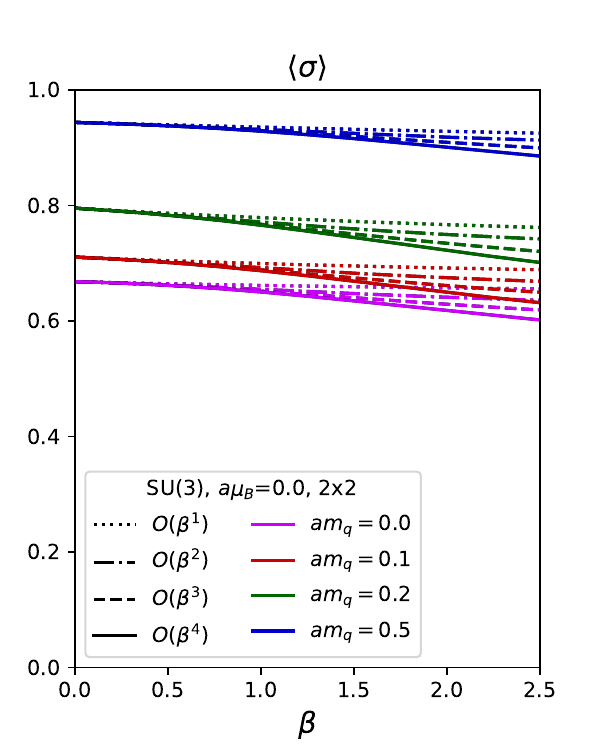}\hspace{-5mm}
 \includegraphics[page=5,width=0.27\textwidth]{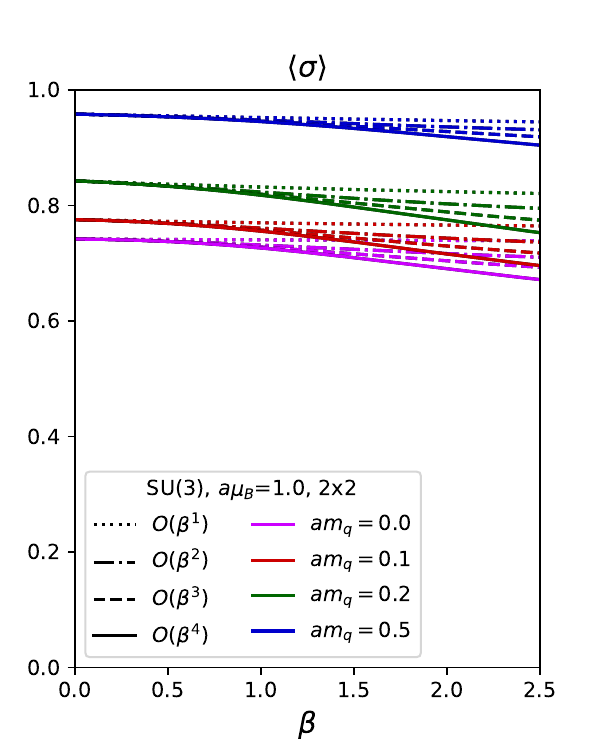}
 }
\centerline{(a)\hspace{40mm}(b)\hspace{48mm}(c)\hspace{40mm}(d)}
\caption{Sign problem on a $2\times 2$ lattice as a function of $\beta$ for various quark masses. (a) Fermionic sign $\langle\sigma_{f}\rangle$ for  $\SU(2)$ which has no sign problem in the strong coupling limit and only a very mild sign problem for finite $\beta$.
(b) Fermionic sign $\langle\sigma_{f}\rangle$ for $\SU(3)$: it remains almost constant with $\beta$ as compared to the strong coupling limit (where the fermionic sign is only due to baryon worldlines). (c) Average sign $\langle\sigma\rangle$ for $\SU(3)$. It includes the sign fluctuations within the tensor network. (d) Same as in (c) considering a non-zero value of the baryon chemical potential $\mu_{B}$.
}
\label{SIGN}
\end{figure*}

Another relevant information we can extract from the exact enumeration concerns the magnitude of the sign problem. A measure of its severity is given by the average sign $\sigma$. It is defined as the ratio of the full ($\mathcal{Z}$) and the so-called phase quenched ($\mathcal{Z}^{p.q.}$) partition function:
\begin{align}
\label{sign}
\langle\sigma\rangle = \frac{\mathcal{Z}}{\mathcal{Z}^{p.q.}}.
\end{align}
The latter is obtained by taking the norm of each statistical weight in $\mathcal{Z}$. In our case, the partition function is a sum of real quantities, hence the norm is just the absolute value. From the definition it follows that $\langle\sigma\rangle \leq 1$ and the equality holds if there is no sign problem. As we want to compare the sign problem in the dual representation with and without the \rhonames{} as an additional degree of freedom, we need to employ two different definitions for the phase quenched system. In the first case, we need to set the fermionic sign $\sigma_{f}=1$ and take the absolute value of each tensor element
\begin{equation}
T_{x}^{\rho^{x}_{-d}\cdots\rho^{x}_{d}} \rightarrow |T_{x}^{\rho^{x}_{-d}\cdots\rho^{x}_{d}}|,
\end{equation}
while in the second case it suffices to set $\sigma_{f}=1$ as a configuration is now determined by the contracted tensor network. The two resulting average signs are respectively $\langle\sigma\rangle = \langle \sigma_{f}\,\sigma_{\rho}\rangle$ and $\langle\sigma_{f}\rangle$. In Fig.~\ref{SIGN} they are shown in the most relevant cases of $\SU(2)$ and $\SU(3)$ as a function of the truncation order and in the $\SU(3)$ case at non-zero baryon chemical potential as well. In the $\SU(2)$ case, the fermionic sign does not play a role on a $2\times 2$ lattice as the allowed loop geometries have $\sigma_{f}=1$ and the only source of negative signs is due to the tensor network. In Fig.~\ref{SIGN} (a), this is shown for various quark masses and for different truncations up to $\mathcal{O}(\beta^{4})$. The trend corresponds to a mild deterioration of the sign as $\beta$ and the truncation order is increased. This deterioration is not dramatic and corresponds to a fall in $\langle\sigma\rangle$ of about $10\%$ at $\beta\approx 2$. In the case of $\SU(3)$, the fermionic sign $\sigma_{f}$ is not positive (Fig.~\ref{SIGN} (b)) but remains almost constant as a function of $\beta$ and truncation order. When considering the sign $\langle \sigma\rangle$, a trend similar to the $\SU(2)$ case shows up (Fig.~\ref{SIGN} (c)): the sign in this case remains almost constant for $\beta\leq 1$ where it starts to get worse as a function of the truncation order. When a non-zero baryon chemical potential is considered (Fig.\ref{SIGN} (d)), this behaviour does not change. \\

\onecolumngrid
\twocolumngrid

\vphantom{abcd}

\vphantom{abcd}

Although our numerical results are preliminary and only based on an exact enumeration of the partition function on a $2\times 2$ lattice, we highlight some of the findings that could extend to larger volumes and higher dimensions. First of all, the comparison with the HMC simulations shows that our method provides the correct Boltzmann weights for the dual configurations as the strong coupling branch up to $\beta\leq 0.5$ is well described by the polynomials $\mathcal{P}\left(\beta,\hat{m}_q,z_{q}\right)$ for different gauge groups and quark masses. This is non-trivial: the number of configurations considered already on a small volume is very large (Tab.~\ref{Tab1}) and the computation very sensitive to the exact evaluation of the tensor elements $T_{x}^{\rho}$ and of the fermionic sign $\sigma_{f}$. The evaluation of the Boltzmann weights away from the strong coupling limit is thus under control and can be used in Monte Carlo simulations if the truncation order $\mathcal{O}(\beta^{n})$ is not too large. We considered two main strategies in view of Monte Carlo simulations: the bubble decomposition Eq.~(\ref{Bubble_decomposition}), and sampling an enlarged configuration space that includes the \rhonames{}.
While we are not yet in position to draw general conclusions on the sign problem, when comparing it with the behaviour in a permutation based dualization~\cite{Gagliardi2018}, the improvement is drastic. Hence, resumming the permutations as in Eq.~(\ref{I-integral_expansion}), effectively reduces the sign problem from the Weingarten functions. \\

\section{Conclusion and discussion}
\label{V}
In this work we proposed a new strategy for the evaluation of higher order contributions in the strong coupling expansion of lattice QCD with staggered fermion discretization.
The dual representation in terms of local tensorial weights improves on the sign problem as compared to evaluations in a Weingarten function basis. 
The color constraints from gauge and Grassmann integration combine to yield admissible configurations that after contracting the tensors are intersecting plaquette surfaces that are either closed or bounded by fermion fluxes. The configuration space is thus a worldline \& worldsheet representation with the additional multi-indices $\rho$ which we called Decoupling Operator Indices and that encode the information about the interplay of the unitary and symmetric group. \\

The prospects of Monte Carlo simulations of lattice QCD at finite density in the strong coupling regime are encouraging: the weights in the partition functions are local, and various strategies to sample the partition function Eq.~(\ref{DualizedPartitionFunction}) are possible.
We will be able to obtain results on the phase diagram in the strong coupling regime beyond $\mathcal{O}(\beta)$. 
One possible way to perform Monte Carlo is via a Worm algorithm based on vertices, as was discussed in the context of the Schwinger model \cite{Wenger2008}. The drawback of this method is that this algorithm slows down drastically with the number of vertices (Tab.~\ref{Tab2}). This limits in practice the maximal order of $\beta$ feasible in 3+1 dimensions.
Another intriguing possibility is to perform local Metropolis updates that could be parallelized. 
We can either sample the multi-indices $\rho$ alongside the occupation numbers (monomer, dimer plaquette and fermion flux) or contract all $\rho$'s on a background of occupation numbers, employing the bubble decomposition discussed in Sec.~\ref{Complexity}. Even when including the higher orders, the sign problem might still be manageable if $\beta$ is not too large. 
For what values of $\beta$ simulations are possible in 3+1 dimensions is only to be seen in practice and will be co-determined by the magnitude of the sign problem, by the numerical cost of evaluating the Boltzmann weights and will also depend crucially on the quark mass. Our representation is also valid for pure Yang Mills theory, which is sign problem free after contracting the tensor network.\\

A finite chemical potential does not introduce an additional sign problem as the zero-density Boltzmann weights get multiplied only by positive factors. Moreover, at fixed values of $\beta$, the sign problem becomes milder for large enough temperatures and/or densities: the worldline configurations contributing to the fermionic sign 
$\sigma_f$ simplify as the quark fluxes are mainly aligned in temporal direction.
A detailed analysis of the sign problem requires, however, large volumes that cannot be obtained via exact enumeration and will be presented in a forthcoming publication. \\

\acknowledgments
We thank Jangho Kim for helpful discussions on fast exact enumeration.
We acknowledge support by the Deutsche Forschungsgemeinschaft (DFG) through the Emmy Noether Program under Grant No. UN 370/1 and through the Grant No.CRC-TR 211 "Strong-interaction matter under extreme conditions". \\

\newpage 
\appendix 
\section{$\SU(N)$ generating functional and $\mathcal{I}$-integrals.}
Eq.~(\ref{I-integral-computation}) for the $\SU(N)$ $\mathcal{I}$-integrals can be derived from the generating functional
\begin{align}
\label{GeneratingFunctional}
Z^{q,p}\left[K,J\right] &= \int_{\SU(N)}DU\left(\Tr [UK]\right)^{qN+p}\,\left(\Tr [U^{\dag}J]\right)^{p},
\end{align}
by taking successive derivatives with respect to the sources $J,K \in GL(N,\mathbb{C})$, according to the following equation:
\begin{widetext}
\begin{align}
\label{Differentiation}
\mathcal{I}^{qN+p,p}\ijkl = \frac{1}{(qN+p)!p!}\frac{\partial^{(qN+2p)} Z^{q,p}\left[K,J\right]}{\partial K_{j_{1}}^{i_{1}}\cdots \partial K_{j_{qN+p}}^{i_{qN+p}}\partial J_{\ell_{1}}^{k_{1}}\cdots\partial J_{\ell_{p}}^{k_{p}}}\bigg|_{J=K=0}.
\end{align}
\end{widetext}
To evaluate $Z^{q,p}[K,J]$, we first convert the integral~[\ref{GeneratingFunctional}] into a $\U(N)$ integral, using
\begin{align}
\label{UN_SUN_correspondence}
\frac{1}{\det K^{q}}&\int_{\SU(N)}DU\left(\Tr [UK]\right)^{qN+p}\,\left(\Tr [U^{\dag}J]\right)^{p} = \nonumber \\
 =&\int_{\U(N)}DU\frac{1}{\det [UK]^{q}}\left(\Tr[UK]\right)^{qN+p}\,\left(\Tr[U^{\dag}J]\right)^{p},
\end{align} 
and assuming for the moment $J,K \in \U(N)$. The equality holds because the last integrand is invariant under multiplication of the $U$ matrix by a complex phase. As a consequence, it gives the same result when integrated using the $\SU(N)$ or the $\U(N)$ Haar measure. Exploiting this trick, we can make use of the $\U(N)$ character expansion to compute the quantity in the r.h.s of Eq.~(\ref{UN_SUN_correspondence}). \\
Thanks to the Schur-Weyl duality~\cite{PPN271034092,weyl1939classical}, power of traces of $\U(N)$ matrices have the following character expansion
\begin{equation}
\label{Equality1}
\left(\Tr U\right)^{n} = \sum\limits_{\substack{\lambda \vdash n \\ \len(\lambda) \leq N}}f_{\lambda}\hat{\chi}^{\lambda}(U),
\end{equation}
where $\hat{\chi}^{\lambda}$ are the $\U(N)$ characters\footnote{Not to be confused with the characters $\chi^{\lambda}$ of the symmetric group.}. Instead, $\det U^{q}$ are irreducible one dimensional representations $\forall q\in\mathbb{Z}$ (so-called determinantal representations). According to a standard group theory result, the tensor product of the irrep. $V_{\lambda}$ with a determinantal representation $V^{q}_{det}$ gives
\begin{equation}
V_{\lambda}\otimes V^{q}_{det} \cong V_{\lambda+ q},
\end{equation}
where $V_{\lambda+q}$ is the $U(N)$ irreducible representation with highest weight: $\{\lambda_{1}+q,\ldots,\lambda_{N}+q\}$. This gives:
\begin{equation}
\label{Equality2}
\hat{\chi}^{\lambda}(U)\, \det(U)^{q} = \hat{\chi}^{\lambda+q}(U).
\end{equation}
Substituting Eqs.~(\ref{Equality1}) ,~(\ref{Equality2}) into the r.h.s. of Eq.~(\ref{UN_SUN_correspondence}) we get:
\begin{widetext}
\begin{align}
\label{Proof}
\frac{Z^{q,p}[K,J]}{\det [K]^{q}}&=
%\int_{\U(N)}DU\frac{1}{\det [UK]^{q}}\left(\Tr[UK]\right)^{qN+p}\left(\Tr[U^{\dag}J]\right)^{p} = 
\int_{\U(N)}DU\!\!\sum_{\substack{\lambda\vdash qN+p \\ \len(\lambda) \leq N}}\sum_{\substack{\lambda'\vdash p \\ \len(\lambda') \leq N}}f_{\lambda} f_{\lambda'}\hat{\chi}^{\lambda-q}(UK)\hat{\chi}^{\lambda'}(U^{\dag}J) =
 \sum_{\substack{\lambda\vdash p \\ \len(\lambda) \leq N}}f_{\lambda+q}\, f_{\lambda}\,\frac{\hat{\chi}^{\lambda}(JK)}{D_{\lambda,N}}  \nonumber \\
&=\frac{(qN+p)!}{p!}{\displaystyle \prod_{i=0}^{N-1}}\frac{i!}{(i+q)!}\sum_{\substack{\lambda\vdash p \\ \len(\lambda) \leq N}}\frac{(f_{\lambda})^{2}}{D_{\lambda,N+p}}\hat{\chi}^{\lambda}(JK) =
\frac{(qN+p)!}{p!^{2}}{\displaystyle \prod_{i=0}^{N-1}}\frac{i!}{(i+q)!}\!\!{\displaystyle\sum_{\substack{\lambda\vdash p \\ \len(\lambda) \leq N}}}\!\!\frac{(f_{\lambda})^{2}}{D_{\lambda,N+q}}\sum_{\substack{\rho\vdash p}}h_{\rho}\hat{\chi}^{\lambda}(\rho)t_{\rho}(JK)  \nonumber \\
&=(qN+p)!{\displaystyle \prod_{i=0}^{N-1}}\frac{i!}{(i+q)!}\sum_{\substack{\rho\vdash p}}h_{\rho}\tWg^{q,p}_{N}(\rho)t_{\rho}(JK),
\end{align}
\end{widetext}
where the second equality follows from the orthogonality of characters, the third from the combinatorial identity
\begin{align}
\frac{f_{\lambda+q}}{D_{\lambda,N}} = {\displaystyle \prod_{i=0}^{N-1}}\frac{i!}{(i+q)!}\frac{f_{\lambda}}{D_{\lambda,N+q}},
\end{align}
valid for $\len(\lambda)\leq N$ and the fourth one from the Frobenius relation (see for instance the Appendix A of~\cite{Drouffe1983}). The last equality is just a rearrangment of terms. The quantities $\tWg_{N}^{q,p}$ are the \textit{Generalized Weingarten functions}:
\begin{align}
\tWg^{q,p}_{N}(\rho)= \frac{1}{(p!)^{2}} \sum_{\substack{\lambda\vdash p\\ \len(\lambda) \leq N}}\frac{(f_{\lambda})^{2}}{D_{\lambda,N+q}}\chi^{\lambda}(\rho).  
\end{align}
They are $S_{p}$ class functions and therefore depend only on the conjugacy class of a given permutation. Conjugacy classes and irreducible representations are in 1-1 correspondence. This is the reason why the Weingarten functions can also have integer partitions as argument. In Eq~(\ref{Proof}), $h_{\rho}$ is the number of permutations within the conjugacy class associated to the partition $\rho$, while $t_{\rho}(JK)$ is a shortcut for
\begin{equation}
t_{\rho}(JK) = {\displaystyle \prod_{i=0}^{\ell(\rho)}}\Tr(JK)^{\rho_{i}}.
\end{equation}
The $\SU(N)$ generating functional is
\begin{widetext}
\begin{align}
\label{GeneratingFunctionalII}
Z^{q,p}[K,J] = (qN+p)!{\displaystyle \prod_{i=0}^{N-1}}\frac{i!}{(i+q)!}\det K^{q}\sum_{\substack{\rho\vdash p}}h_{\rho}\tWg^{q,p}_{N}(\rho)t_{\rho}(JK),
\end{align}
\end{widetext}
and given the polynomial nature of the expression, it can be extended to any $K,J\in GL(N,\mathbb{C})$. In the limits $q=0$, $q=1$ and $p=0$ the known results~\cite{Drouffe1983},\cite{Zuber_2016} and~\cite{Creutz:1978ub} are recovered. \\
Given the expression~(\ref{GeneratingFunctionalII}), the $\mathcal{I}$-integral is obtained by taking derivatives with respect to the sources $K,J$. We do not need to do this explicitly. In fact, it is sufficient to know the result in the cases $p=0$ and $q=0$ and then use Leibnitz Formula for the derivative of a product. Luckily, these two special cases have already been solved respectively by Creutz~\cite{Creutz:1978ub} and by Collins and collaborators in~\cite{Collins1,Collins2}. The two results are
\begin{widetext}
\begin{align}
\label{CollinsCreutz}
\mathcal{I}^{qN,0}\ijkl &=\frac{1}{(qN)!}\frac{\partial^{(qN)}Z^{q,0}[K]}{\partial K_{j_{1},i_{1}}\cdots K_{j_{qN},i_{qN}}}\bigg|_{J=K=0} = \left[{\displaystyle\prod_{i=1}^{N-1}}\frac{i!}{(i+q)!}\right]\sum_{\{\alpha\}}\epsilon^{\otimes q}_{i_{\{\alpha\}}}\epsilon^{\otimes q, j_{\{\alpha\}}}, \nonumber \\
\mathcal{I}^{p,p}\ijkl &=\frac{1}{p!^{2}}\frac{\partial^{(2p)}Z^{0,p}[K,J]}{\partial K_{j_{1},i_{1}}J_{l_{1},k_{1}}\cdots K_{j_{p},i_{p}}J_{l_{p},k_{p}}}\bigg|_{J=K=0} = {\displaystyle \sum_{\pi,\sigma \in S_{p}}}\delta_{i}^{l_{\pi}}\tWg_{N}^{0,p}\left(\pi\circ\sigma^{-1}\right)\delta_{k_{\sigma}}^{j},
\end{align}
\end{widetext}
where $\delta_{i}^{l_{\pi}}$ and $\delta_{k_{\sigma}}^{j}$ are the usual Kronecker deltas where the indices are swapped according to permutations $\pi$ and $\sigma$. The sum in the first line of Eq~(\ref{CollinsCreutz}) runs over all possible ways $\alpha=\{\alpha_{1},\ldots,\alpha_{q}\}$ of partitioning the $qN$ indices into $q$ epsilon tensors.
To get the general $\mathcal{I}$-integral it is sufficient to exploit the fact that the the generating functional~(\ref{GeneratingFunctionalII}) can be decomposed, apart from a trivial combinatorial factor, as a product of $Z^{q,0}$ and a term that resemble the generating functional $Z^{0,p}$. The only difference is in the coefficients $\tWg_{N}^{0,p}$ that must be substituted with $\tWg_{N}^{q,p}$. Therefore, by looking at Eq.~(\ref{Differentiation}), when $qN$ derivatives of $K$ act on the power of the determinant $\det K^{q}$, they will reproduce the result for $I^{qN,0}$. Similarly, when $p$ derivatives of $K$ and $p$ derivatives of $J$ act on the second term, they will reproduce $I^{p,p}$ with the substitution $\tWg_{N}^{0,p}\rightarrow \tWg_{N}^{q,p}$. Any other combination of derivatives gives zero. Making use of the Leibnitz Formula, we can thus write down the expression of the $\mathcal{I}$-integral as
\begin{widetext}
\begin{equation}
\mathcal{I}^{qN+p,p}\ijkl = \left[{\displaystyle \prod_{i=1}^{N-1}}\frac{i!}{(i+q)!}\right]{\displaystyle \sum_{\{\alpha,\beta\}}}{\displaystyle \sum_{\pi,\sigma \in S_{p}}}\epsilon^{\otimes q}_{i_{\{\alpha\}}}\delta_{i_{\{\beta\}}}^{l_{\pi}}\tWg_{N}^{q,p}(\pi\circ\sigma^{-1})\epsilon^{\otimes q, j_{\{\alpha\}}}\delta_{k_{\sigma}}^{j_{\{\beta\}}},
\end{equation}
\end{widetext}
where the leftmost sum now runs over all the ways $(\alpha,\beta)$ of partitioning the $i$, $j$ indices into the Kronecker deltas and into the $q$ epsilon tensors. This "multplicity" stems from the fact that we need to take into account every possible way of acting with the $K$ derivatives, on the determinant and on the traces $\tr_{\rho}(JK)$, and from the Creutz result for $I^{qN,0}$ in Eq.~(\ref{CollinsCreutz}).  \\ 
\label{App1}

\section{(Re-)Derivation of the $\mathcal{O}(\beta)$ partition function}
\begin{center}
\begin{figure*}
\includegraphics[scale=0.15]{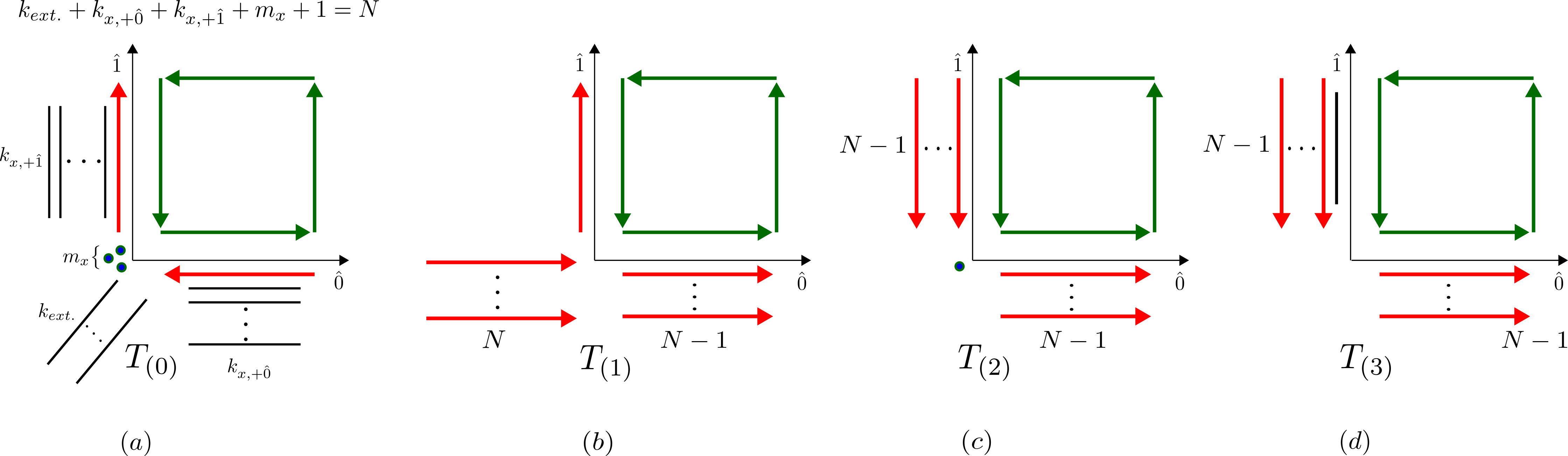}
\caption{The four different types of tensor at a corner of the excited plaquette. $(a)$ The two excited links are occupied by dimers and a single quark flux. It represents the most general $\U(N)$ contribution to the $\mathcal{O}(\beta)$ partition function. (b) An incoming baryon $(f_{\ell}=N)$ split into a $N-1$ quark flux and a single quark flux. (c) An $N-1$ quark flux travels in the same direction of the gauge flux. A dimer or a monomer must be present to satisfy the Grassmann constraint. (d) As in (c) with a dimer superimposed to the $N-1$ quark flux on one of the two excited links.}
\label{Fig8}
\end{figure*}
\end{center}
The partition function~(\ref{DualizedPartitionFunction}) at $\mathcal{O(\beta)}$ can be rewritten in terms of site and link (scalar) weights. This is done considering all possible tensors $T_{x}$ at a corner of the excited plaquette, showing that they can be reduced to scalar objects. In this limiting case, we can perform all steps analytically so as to recover the partition function obtained in~\cite{deForcrand2014}. For general $\SU(N)$ there are two types of tensors: those corresponding to genuine $\SU(N)$ contributions (Fig.~\ref{Fig8} (b)-(d)) and the ones that correspond to dimers with a single quark flux oppositely oriented with respect to the plaquette (Fig.~\ref{Fig8} (a)). The latter is a pure $\U(N)$ contribution as the associated $\mathcal{I}$-integrals have $q=0$. Let us consider first the second case. Even though the corresponding tensors can be quite large, the same decoupling present at strong coupling, holds in this case. Proceeding in a similar fashion as in Eqs.~(\ref{Sc_derivationI}),~(\ref{Sc_derivationII}),~(\ref{Strong_coupling_part_function}) and with reference to Fig.~\ref{Fig8} (a), one gets
\begin{widetext}
\begin{align}
T^{\rho^{x}_{-d}\cdots\rho^{x}_{d}}_{(0)} = N!{\displaystyle \prod_{\mu \in s.c.}}\frac{1}{\sqrt{D_{\lambda_{\mu},N}}}\delta_{\lambda_{\mu},\;\tiny\Yvcentermath4\yng(1,1,1)\;\big\rbrace{\;k_{x,\mu}^{s.c.}}}\,\frac{1}{\sqrt{D_{\lambda^{\hat{0}},N}}}\delta_{\lambda^{\hat{0}},\;\tiny\Yvcentermath4\yng(1,1,1)\;\big\rbrace{\;k_{x,+\hat{0}}+1}}\,\frac{1}{\sqrt{D_{\lambda^{\hat{1}},N}}}\delta_{\lambda^{\hat{1}},\;\tiny\Yvcentermath4\yng(1,1,1)\;\big\rbrace{\;k_{x,+\hat{1}}+1}},
\end{align}
\end{widetext}
where the first product runs over the external (strong coupling) dimers. As in the strong coupling limit, only one element of the tensor is non-zero. The modification of the dimer weight is obtained as in Eq.~(\ref{Strong_coupling_part_function})
\begin{align}
\label{DimerOB}
&w_{\ell}\left(k_{\ell},f_{\ell} =\pm 1 \right) = \frac{1}{k_{\ell}!(k_{\ell}+1)!}\,\frac{1}{D_{\,\,\tiny\Yvcentermath4\yng(1,1,1)\;\big\rbrace{\;k_{\ell}+1},N}}= \nonumber \\
&= \frac{1}{k_{\ell}!(k_{\ell}+1)!}\,\frac{(k_{\ell}+1)!}{N(N-1)\cdots (N-k_{\ell})}= \nonumber \\
&= \frac{(N-k_{\ell}-1)!}{N!k_{\ell}!},
\end{align}
and gives the correct link weight to be used when a dimer belongs to an excited link. The genuine $\SU(N)$ configurations are instead of three types:
\begin{enumerate}[1)]
\item An incoming (strong coupling) baryon splits, at a corner of the plaquette, into a single quark flux and a $N-1$ quark flux. Equivalently, a single quark flux and a $N-1$ quark flux can recombine to form an outcoming (strong coupling) baryon (Fig.~\ref{Fig8} (b)).
\item An incoming $N-1$ quark flux exits the site following the gauge flux induced by the plaquette. A monomer or an external dimer is also present in order to fulfill the Grassmann constraint (Fig.~\ref{Fig8} (c)).
\item As in $2)$ with the external dimer or monomer replaced by a dimer on one of the two excited links (Fig.~\ref{Fig8} (d)).
\end{enumerate}
The first two types of configurations are somewhat trivial as the associated tensors have size one. There is in fact only one \rhoname{} associated to the external legs of the two tensors. Their values can be readily computed:
\begin{equation}
\label{Tensor12}
T_{(1)} =  \frac{N!}{\sqrt{N}} \qquad\quad\qquad T_{(2)}= (N-1)!.
\end{equation}
In the case of configurations of type 3), the associated tensor has size $2\times 1$. There are in fact two \rhonames{} in direction $+\hat{1}$, where a dimer is superimposed to a $N-1$ quark flux. This tensor is given by:
\begin{equation}
T_{(3)}^{1,1} = \frac{N!}{\sqrt{N+1}} \quad\quad\qquad T_{(3)}^{1,2} = \frac{N!}{\sqrt{N(N+1)}}.
\end{equation}
To remove this multiplicity, it is sufficient to notice that a link carrying a dimer plus a $N-1$ quark flux can only recombine with a $N-1$ quark flux from another direction. The latter involves an $\mathcal{I}$-integral made up of a single Decoupling Operator. Therefore, we can perform a resummation of the two \rhonames{} by considering the following modified "tensor" of size $1$
\begin{equation}
\label{Tensor3}
\tilde{T}_{(3)} = \sqrt{\left(T_{(3)}^{1,1}\right)^{2} + \left(T_{(3)}^{1,2}\right)^{2}} = \frac{N!}{\sqrt{N}},
\end{equation}\\
and all tensors have been thus reduced to scalar quantities. It is easy to check that the modified dimer weights (Eq.~(\ref{DimerOB})) and the values of $T_{(1)},T_{(2)},\tilde{T}_{(3)}$ together with the usual combinatorial factors from the Taylor expansion, are recovered by defining the following link and site weights at the boundary of the excited plaquette:
\begin{enumerate}[1)]
\item To each $N-1$ quark flux associate a link weight: $\frac{1}{N!(N-1)!}$.
\item To each $N-1$ quark flux superimposed to a dimer associate a link weight: $\frac{(N-1)!}{N!}=\frac{1}{N}$
\item For $k_{\ell}\in\{0,\ldots,N-1\}$ dimers and a single quark flux associate $w_{\ell}(d_{\ell},f_{\ell}=\pm 1)$. 
\item To each site corresponding to a $\U(N)$ configuration (Fig.~\ref{Fig8}(a)) associate the usual site weight:  $N!/m_{x}!$.
\item To each site corresponding to a $\SU(N)$ configurations, associate a factor $N!$ if there are no external (strong coupling) dimers or baryons and if $m_{x}=0$. Associate a factor $N!(N-1)!$ if there are external dimers or if $m_{x}=1$. Finally, associate $N!\sqrt{N}$ if there is an external baryon. 
\end{enumerate}
The rules (1)-(5) together with the usual strong coupling weights define the $\mathcal{O}(\beta)$ partition function~\cite{deForcrand2014}. Beyond this order, it is not possible to reduce the tensor network to a product of scalar link and site weights depending only on $\{n_{p},\bar{n}_{p},k_{\ell},f_{\ell},m_{x}\}$.  
\label{App2}

\bibliography{StaggeredGaugeCorr.bib} 
\bibliographystyle{apsrev4-1}

\end{document}